\documentclass[aps,prx,twocolumn,notitlepage,showpacs,amsmath,amstex,amssymb,citeautoscript,longbibliography]{revtex4-2}
\pdfoutput=1
\usepackage{natbib}
\usepackage[english]{babel}
\usepackage{letltxmacro}
\usepackage{latexsym}
\usepackage{floatrow}
\LetLtxMacro{\ORIGselectlanguage}{\selectlanguage}
\makeatletter
\DeclareRobustCommand{\selectlanguage}[1]{%
  \@ifundefined{alias@\string#1}
    {\ORIGselectlanguage{#1}}
    {\begingroup\edef\x{\endgroup
       \noexpand\ORIGselectlanguage{\@nameuse{alias@#1}}}\x}%
}
\newcommand{\definelanguagealias}[2]{%
  \@namedef{alias@#1}{#2}%
}
\makeatother
\definelanguagealias{en}{english}
\definelanguagealias{English}{english}
\usepackage{graphicx}
\usepackage{amsmath}
\usepackage{braket}
\usepackage{amsfonts}
\usepackage{amssymb}
\usepackage{bm}
\usepackage{color}
\usepackage{soul}
\usepackage{wasysym}
\usepackage{dsfont}
\usepackage{bbold}
\usepackage{bbm}

\usepackage{hyperref}
\hypersetup{
    bookmarks=false,         
    unicode=false,          
    pdftoolbar=false,        
    pdfmenubar=true,        
    pdffitwindow=false,     
    pdfstartview={FitH},    
    pdftitle={Entanglement view of dynamical quantum phase transitions},    
    pdfauthor={Stefano De Nicola, Alexios A. Michailidis, Maksym Serbyn},     
    pdfsubject={},   
    pdfcreator={},   
    pdfproducer={}, 
    pdfkeywords={MPS, DQPT, entanglement}, 
    pdfnewwindow=true,      
    colorlinks=true,       
    linkcolor=black,          
    citecolor=blue,        
    filecolor=magenta,      
    urlcolor=blue           
}

\newcommand{\dd}{\mathrm{d}}

\begin{document}
\title{Entanglement view of dynamical quantum phase transitions}
\author{Stefano De Nicola}
\author{Alexios A. Michailidis}
\author{Maksym Serbyn}
\affiliation{IST Austria, Am Campus 1, 3400 Klosterneuburg, Austria}
\date{\today}
\begin{abstract}
The analogy between an equilibrium partition function and the return probability in many-body unitary dynamics has led to the concept of dynamical quantum phase transition~(DQPT). DQPTs are defined by non-analyticities in the return amplitude and are present in many models. In some cases DQPTs can be related to equilibrium concepts such as order parameters, yet their universal description is an open question. In this work we provide first steps towards a classification of DQPTs by using a matrix product state description of unitary dynamics in the thermodynamic limit. This allows us to distinguish the two limiting cases of \emph{precession} and \emph{entanglement} DQPTs, which are illustrated using an analytical description in the quantum Ising model. While precession DQPTs are characterized by a large entanglement gap and are semiclassical in their nature, entanglement DQPTs occur near avoided crossings in the entanglement spectrum and can be distinguished by a complex pattern of non-local correlations. We demonstrate the existence of precession and entanglement DQPTs beyond Ising model, discuss observables that can distinguish them and relate their interplay to complex DQPT phenomenology.
\end{abstract}

\maketitle
\emph{Introduction.---}The rapid development of different quantum simulation platforms \cite{reviewColdAtoms2015,Gross2017} fuels the exploration of new non-equilibrium  phenomena that can be probed in isolated interacting quantum systems. Due to experimental limitations, phenomena observable at short times in quantum quenches are of particular interest. Dynamical quantum phase transitions (DQPTs) have recently emerged as an interesting phenomenon within this regime~\cite{heyl2013,heyl2018}. 
Since the early work of~\textcite{heyl2013}, who introduced the notion of DQPT considering the quantum Ising model, DQPTs have attracted great interest \cite{KarraschSchuricht2013,vajnaDora2014,andraschkoSirker2014,
canovi2014,CanoviPRB2014,Torlai2014,heyl2014,heyl2015,vajnaDora2015,
Sharma2015,schmittKehrein2015,Halimeh2017,weidinger2017,Karrasch2017,Homrighausen2017,
zunkovic2018,schmittHeyl2018,Trapin2018,Gurarie2019,DeNicola2019,
Huang2019,Lacki2019,jafari2019,heyl2019}. Moreover, they were experimentally observed in trapped ion quantum simulators~\cite{jurcevic2017}, superconducting qubits~\cite{Guo2019} and other platforms~\cite{Flaschner2018,Tian2018,Wang2019,Xu2020}.

In the framework of DQPTs, one considers quantum quenches from an initial state $\ket{\psi_{0}}$ and monitors the normalized logarithm of the return probability in the process of unitary evolution under a Hamiltonian $H$,
\begin{equation}\label{eq:fid}
f(t) = -\lim_{L\rightarrow \infty}\frac{1}{L}\log{|\bra{\psi_{0}}e^{-i H t}\ket{\psi_{0}} |^2}, 
\end{equation}
where we restrict to one dimensional cases and denote system size as $L$.
This quantity is identified as the non-equilibrium analogue of the free energy density, with DQPTs corresponding to non-analyticities in the behavior of $f(t)$ at early times~\cite{heyl2013}. 
However, $f(t)$ corresponds to the free energy at complex temperature, and a precise relation between the behavior of $f(t)$ and the equilibrium phase diagram was not established~\cite{heyl2019}. 
Phenomenologically, quenches from a state $\ket{\psi_{0}}$ that realizes a different phase compared to the ground state of $H$ often give rise to DQPTs \cite{heyl2013,KarraschSchuricht2013,Torlai2014,Karrasch2017,heyl2018}; however, there are exceptions from this rule \cite{andraschkoSirker2014,vajnaDora2014,Sharma2015,schmittKehrein2015,jafari2019,heyl2019}. 

In order to connect DQPTs to equilibrium concepts, such as order parameters, the behavior of local observables was explored~\cite{Trapin2018,Lacki2019,Rylands2020}. A direct correspondence was established for systems with broken symmetries, involving a generalized notion of DQPTs~\cite{heyl2014,weidinger2017,zunkovic2018,feldmeierPollmannKnap2019}. Recently, local string observables capable of revealing DQPTs were introduced~\cite{halimeh2020local,bandyopadhyay2020observing}. 
However, the general relation between DQPTs and local expectation values remains elusive.
Connections to the entanglement entropy were also explored: DQPTs may correspond to regions of rapid growth~\cite{jurcevic2017} or peaks~\cite{schmittHeyl2018} in the entanglement entropy, and, for certain quenches in integrable models, they occur at crossings in the entanglement spectrum~\cite{CanoviPRB2014,Torlai2014,Surace2020}. Nonetheless, the underlying mechanism and the conditions under which DQPTs may be related to entanglement signatures are not clearly understood. Thus, in spite of many advancements, the rich phenomenology of DQPTs and their relation to other physical quantities still call for a more general understanding~\cite{heyl2019}.

In this manuscript, we utilize the matrix product state (MPS)~\cite{Paeckel2019} language for DQPTs that was applied in numerical studies~\cite{KarraschSchuricht2013,Torlai2014,heyl2014,Sharma2015,Karrasch2017,Lacki2019}. We show that in the low-entanglement regime it is possible to distinguish between \textit{precession} and \textit{entanglement} DQPTs, which correspond to different physics, as highlighted by analytical MPS ans\"atze. We illustrate the existence of entanglement and precession DQPTs in different models, discuss ways to distinguish them experimentally and suggest how their interplay may lead to the rich phenomenology reported in the literature.  

\begin{figure*}[t]
\begin{center}
\includegraphics[width=\columnwidth]{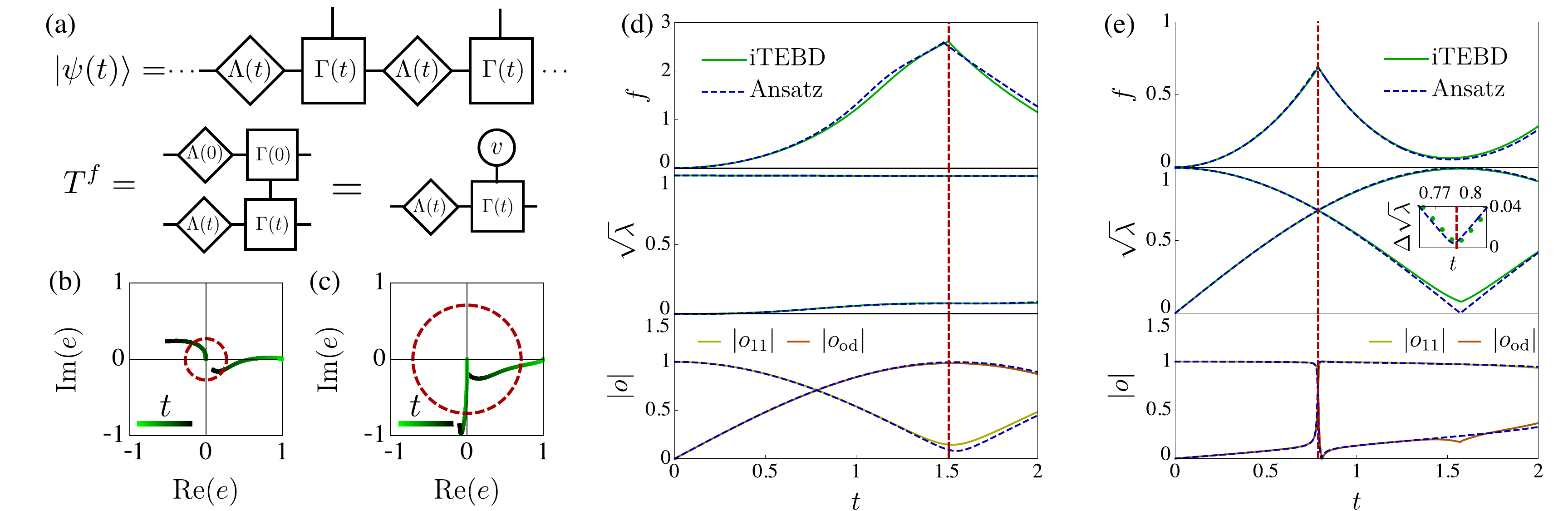}\\
\end{center}
\caption{\label{fig:intro}
(a) Representation of the fidelity density transfer matrix $T^f$ obtained from the iMPS canonical form when the initial state is a product state. 
The evolution of the two leading eigenvalues of $T^f$ in the complex plane, the fidelity density, the entanglement spectrum and the overlaps are illustrated for a \emph{pDQPT} in panels (b) and (d) and for an \emph{eDQPT} in (c) and (e).  Red circles in panels (b)-(c) correspond to times where DQPTs occur. Panels (d)-(e) compare fidelity density, entanglement spectrum and overlaps. Solid lines show iTEBD data obtained with $\chi\leq 8$ (truncating $\sqrt{\lambda}_i<10^{-9}$), and dashed lines correspond to the analytical ans\"atze; quench parameters are listed in the main text. 
%
 %
}
\end{figure*} 

\emph{MPS description of DQPTs.---}
DQPTs are typically studied at short times for quenches from area-law entangled initial states. In this regime, the time-evolved state $\ket{\psi(t)} = e^{-i H t}\ket{\psi_{0}}$ has area-law  entanglement due to Lieb-Robinson bounds~\cite{Lieb1972,Eisert2006} and admits an MPS description~\cite{Paeckel2019}. For translation-invariant initial states (possibly with a finite-size unit cell), infinite MPS (iMPS)~\cite{Vidal2006} provides an efficient representation of $\ket{\psi(t)}$. In Fig.~\ref{fig:intro}(a) we show a iMPS in the canonical form~\cite{Vidal2006,Orus2008}, where the standard building block of MPS, the tensor $A^\sigma_{ij}(t)$, is decomposed as $A^\sigma_{ij}(t) =  \Lambda_{ii}(t) \Gamma^\sigma_{ij}(t)  $. Here $\sigma = \uparrow, \downarrow$ is the physical and $i,j=1,\ldots, \chi$ are bond indices. The diagonal matrix $\Lambda_{ii}(t) = \sqrt{\lambda}_i$ contains the ordered,  $\lambda_i>\lambda_{i-1}$, singular values of the Schmidt decomposition across a bond. $\lambda_i$ determine the entanglement spectrum, so that the bipartite entanglement entropy is $S=-\sum_i \lambda_i \log \lambda_i$. 
The tensor $\Gamma^\sigma(t)$ carries a physical index and together with $\Lambda$ satisfies a set of canonical conditions, 
$\sum_{ij\sigma} \Lambda_{ij}^2 \Gamma^\sigma_{jk} \Gamma^{\sigma *}_{il}=  \sum_{ij\sigma} \Lambda_{ij}^2 \Gamma^\sigma_{kj} \Gamma^{\sigma *}_{li}  = \delta_{kl}$~\cite{
Orus2008}. 

Using the iMPS representation, the fidelity density is expressed directly in thermodynamic limit via the spectrum of the fidelity transfer matrix, ${T^f}(t)$, $\{e_{i}\}$, as \cite{andraschkoSirker2014,piroli2018} 
\begin{align}\label{eq:fidelityMPS}
f(t)=- 2 \log \text{max}\left(\{|e_{i}|\}\right).
\end{align}
The transfer matrix ${T^f}(t)$ is defined in Fig.~\ref{fig:intro}(a) as a contraction of the time-evolved MPS tensor with its conjugate at $t=0$. Thus, $T^f(0)$ coincides with the conventional transfer matrix and has $|e_1|=1$ and all other $|e_i|<1$, as follows from normalization of $\ket{\psi_0}$. At later times the eigenvalues of ${T^f}(t)$ perform complicated evolution in the complex plane. As illustrated in Fig.~\ref{fig:intro}(b)-(c), singularities in $f(t)$ emerge from the initially subleading eigenvalue, $e_2$, surpassing in magnitude the largest one.

\emph{Two limiting cases of DQPTs.---}To distinguish between different physical mechanisms that drive the crossing between transfer matrix eigenvalues, we use the canonical form of the MPS tensor and focus on the case when the initial state $\ket{\psi_{0}} = \otimes_i \ket{v}_i$ is a product state. The contraction of the time-evolved MPS with the product state does not affect bond indices, see Fig.~\ref{fig:intro}(a), resulting in 
\begin{equation}\label{eq:Tf}
T^f(t) =
\left(
\begin{matrix}
\sqrt{\lambda}_1&0\\
0&\sqrt{\lambda}_2\\
\end{matrix}
\right)
\left(
\begin{matrix}
o_{11}&o_{12}\\
o_{21}&o_{22}\\
\end{matrix}
\right)
,\ o_{ij} = \sum_\sigma (v^\sigma)^* \Gamma^\sigma_{ij},
\end{equation}
where we retained the leading $2\times 2$ part of the MPS virtual space, corresponding to the two largest singular values. For initial product states, the elements of the overlap matrix $o$ are obtained via contraction of the tensor $\Gamma^\sigma_{ij}$ with the single-site spinor wave function~$v^\sigma$. 

Equations~(\ref{eq:fidelityMPS})-(\ref{eq:Tf}) single out the contribution of the entanglement spectrum, encoded in the diagonal of the matrix $\Lambda$, to the transfer matrix ${T^f}(t)$ and DQPTs. When the entanglement spectrum features a large gap, $\lambda_1\gg \lambda_2$, the switch in magnitude between eigenvalues of $T^f$ is necessarily driven by the evolution of the overlap matrix. This is a \emph{precession} DQPT (pDQPT) that is of semiclassical nature, as we explain below. In the opposite limit, when the two leading singular values $\lambda_1$ and $\lambda_2$ exhibit an avoided crossing, the system features  entanglement of order $\ln 2$. DQPTs happening near such points are dubbed \emph{entanglement} DQPTs (eDQPTs). We illustrate these two limits of DQPTs in the quantum Ising model using analytical MPS ans\"atze. 
  
\emph{Precession DQPTs in the Ising model.---}In order to illustrate pDQPTs, we study the dynamics  under the transverse and longitudinal-field Ising model
\begin{equation}\label{eq:Hising}
H= \sum_{i}\left[J\sigma_{i}^{z}\sigma_{i+1}^{z}+h_{x}\sigma_{i}^{x}+h_{z}\sigma_{i}^{z}\right].
\end{equation} 
The initial state $\ket{\psi_0}=\otimes_i \ket{\downarrow}_i$ is the ground state of the Hamiltonian~(\ref{eq:Hising}) in the ferromagnetic phase, $J\to -\infty$, $h_z>0$. The evolution is performed with $J =0.1$, $h_x =1$, $h_z =0.15$, so that single-spin terms are dominant. 

The top panel of Fig.~\ref{fig:intro}(d) shows the fidelity density calculated using infinite time-evolving block decimation (iTEBD)~\cite{Vidal2006}. It exhibits a cusp at $t\approx 1.5$, signaling a DQPT. By bringing the MPS to the canonical form we extract the entanglement and overlap contributions. The middle plot shows the evolution of the two leading singular values, which remain very well separated at the time when the DQPT occurs. At the same time, $|o_{11}|$ exhibits a minimum near the DQPT, while the off-diagonal component $|o_{12}|=|o_{21}| \equiv |o_{\text{od}}|$ shows a clear maximum. Thus, the overlap matrix is predominantly responsible for the switch of the transfer matrix eigenvalues, providing a prototypical example of pDQPT. 

\emph{Analytical pDQPT ansatz.---}The precession nature of pDQPTs can be illustrated by analytically constructing a suitable $\chi=2$ MPS ansatz. In the limit when $J\ll h_x, h_z$ we split the Hamiltonian~(\ref{eq:Hising}) into an interacting part $V=J\sum_i \sigma_{i}^{z}\sigma_{i+1}^{z}$ and a free-precessing part  $H_0= \sum_i[h_{x}\sigma_{i}^{x}+h_{z}\sigma_{i}^{z}]$ that contains only single-spin terms. Then, we move to the rotating frame with respect to $H_0$, rewriting the time evolution as
$\ket{\psi(t)}  = e^{-i  H_0 t} \mathrm{T} e^{-i\int_0^t {\tilde V}(t^\prime)  \mathrm{d}t^\prime} \ket{\psi_0}$.
The interaction term in the rotating frame reads:
$
{\tilde V}(t) = e^{i t H_0} V e^{-i t H_0} =  \sum_{i}\sum_{\alpha,\beta} s_\alpha(t) s_\beta(t) \sigma^\alpha_i\sigma^\beta_{i+1},
$
where $\alpha, \beta \in \{x,y,z\}$, the time-dependent coefficients are $s_x(t) = 2 h_xh_z\sin ^2(h t)/h^2$, $s_y(t)= h_x\sin (2 h t) /h$, $s_z(t) =[ h_x^2 \cos (2 h t)+h_z^2]/h^2$, and $h = \sqrt{h_x^2+h_z^2} $ is the magnitude of the applied field. Finally, we exploit the slow initial buildup of entanglement along the $z$ axis to replace the $\sigma^x$ and $\sigma^z$ operators in ${\tilde V}(t)$ by their expectation values under free precession, $-s_x$ and $-s_z$ respectively. This allows us to approximately write $\tilde V(t)$ as a matrix product operator (MPO) of $\chi=2$~\cite{SOM} \cite{Crosswhite2008} \cite{mussardo2009}. Acting by this MPO on the initial $\ket{\downarrow}$-product state gives an MPS ansatz for $\ket{\psi(t)}$. Bringing this ansatz to canonical form~\cite{SOM}, we obtain the singular values as $\sqrt{\lambda}_1 = |\cos [J a(t)]|$, $\sqrt{\lambda}_2 = | \sin [J a(t)]|$, where $a(t) =h_x^2 [4 h t-\sin (4 h t)]/8 h^3$. The middle panel of Fig.~\ref{fig:intro}(d) reveals an excellent agreement between our analytical results and iTEBD predictions for the singular values. 
The $\Gamma$ matrix in the canonical form reads:
\begin{equation}\label{eq:Gamma-p}
\Gamma(t) =   e^{-it ( h_x \sigma^x+ h_z\sigma^z) } e^{-i J b(t)\sigma^y}\left(\begin{matrix} \ket{\downarrow} & \ket{\uparrow} \\
i \ket{\uparrow}  & -i \ket{\downarrow}  \end{matrix} \right) \bar{\Lambda},
\end{equation}
where $\bar{\Lambda}=\text{diag}(\, \text{sign}[ \cos (J a(t))],\, \text{sign} [\sin (J  a(t))] )$ and $b(t)  = h_x [h_x^2 \cos (6 h t)+3 \left(h^2+3 h_z^2\right) \cos (2 h t)- 4 \left(h^2+2 h_z^2\right)  ]/12 h^4$~\cite{SOM}.
The matrix of overlaps $o$ is obtained by contracting all entries of $\Gamma(t)$  with the $\bra{\downarrow}$ state on the left. The behavior of $o_{11}$ and $o_{\text{od}}$ obtained from (\ref{eq:Gamma-p}) agrees with numerically exact iTEBD results, Fig.~\ref{fig:intro}(d). Since $\lambda_1\gg \lambda_2$ within the range of considered times, the precession of spins in $\Gamma(t)$ induced by exponentials of Pauli matrices plays the main role in driving the pDQPT. 

The dominant component of the MPS corresponds to the top diagonal entry in Eq.~(\ref{eq:Gamma-p}), and it coincides with the initial state $\ket{\downarrow}$ at $t=0$. The off-diagonal entries in Eq.~(\ref{eq:Gamma-p}) give subleading contributions suppressed by powers of $\sqrt{\lambda_2/\lambda_1}$, as follows from Eq.~(\ref{eq:Tf}). However, as both the dominant component $\ket{\downarrow}$ and its correction $\ket{\uparrow}$ precess, see Eq.~(\ref{eq:Gamma-p}) and \cite{SOM}, the overlap of the dominant contribution decreases while the subleading state rotates closer to the $\ket{\downarrow}$ state. A pDQPT occurs when the formerly subleading contribution becomes important enough to flip the magnitude of the eigenvalues of $T^f$, which happens when $|o_{11}/o_{\text{od}}|\sim \sqrt{\lambda_2/\lambda_1} \ll 1$. The DQPT is then closely associated with the minimum of $o_{11}$, with corrections given by off-diagonal terms; see Fig.~\ref{fig:intro}(d). Note that, although free precession dominates the dynamics for the present quench, a minimal $\chi=2$ is required to capture DQPTs due to Eq.~(\ref{eq:fidelityMPS}), reflecting the quantum nature of such phenomena.

\emph{Entanglement DQPTs in the Ising model.---}We consider a quench from the initial state $\ket{\psi_0}=\otimes_i \ket{\rightarrow}_i$, corresponding to the free paramagnet ground state of~(\ref{eq:Hising}) for $h_x\rightarrow-\infty$. The dynamics is governed by the Ising Hamiltonian with $J =1$, $h_x =0.1 $, $h_z =0.15$. Figure~\ref{fig:intro}(e) shows that a DQPT happens near an avoided crossing in the entanglement spectrum. The overlaps $|o_{11}|$ and $|o_{\text{od}}|$ also display the evolution characteristic of an avoided crossing. This provides an example of eDQPT. 

\emph{Analytical eDQPT ansatz.---}The smallness of all but the first two singular values for this quench allows us to analytically construct a $\chi = 2$ MPS ansatz describing eDQPTs, which agrees well with numerically exact iTEBD. To this end, we approximate the time-evolution operator by a second-order Trotter decomposition, splitting the Hamiltonian into a single-spin term, $H_0$, and a two-spin term $V$. The decomposition reads: $e^{-iHt} \approx e^{-iH_0 t/2}e^{-iVt} e^{-iH_0 t/2}$, where $e^{-iVt}$ admits an exact MPO representation with $\chi=2$, see~\cite{SOM}. Applying the resulting MPO to the initial state we obtain the analytical MPS ansatz 
\begin{align}\label{eq:ansatz_eDQPT}
A(t)= \left(\begin{matrix}e^{-iJt} c_{\uparrow}(t)  \ket{\uparrow(t)}& e^{i Jt} c_{\uparrow}(t) \ket{\uparrow (t)}  \\ e^{iJt } c_{\downarrow}(t) \ket{\downarrow (t)}  & e^{-iJt} c_{\downarrow}(t) \ket{\downarrow (t)}  \end{matrix} \right),
\end{align}
where $\ket{\uparrow(t)} = \exp[ - i t (h_{x}\sigma^{x}+h_{z}\sigma^{z})/2)]\ket{\uparrow} $ and $c_{\uparrow}(t) = \langle \uparrow \ket{ \rightarrow (t)}$, and likewise for $\downarrow$.

Casting the ansatz (\ref{eq:ansatz_eDQPT}) in canonical form yields the tensor $\Gamma$, which generally has a complicated expression but can be simplified in certain limits~\cite{SOM},  and the entanglement spectrum $\lambda_{1,2} = [4 \pm  \sqrt{f(t)+13}]/8 $, whose avoided crossings are expected to drive the DQPT. Here $f(t) = 4 \cos (4 \theta(t) )-\cos (8 \theta(t) )+8 \sin ^4(2 \theta(t) ) \cos (4 J t)$ is expressed in terms of a time-dependent angle
$ 2\cos^2\theta(t)= 1+[1-\cos(h t)] h_x h_z/h^2$. The special cases when either $h_x$ or $h_z$ vanishes correspond to classical~\cite{heyl2015,Trapin2018} or integrable~\cite{CanoviPRB2014,Torlai2014,Surace2020,Calabrese2012,heyl2013} Ising models, discussed in~\cite{SOM}.

In the generic case with $h_x,h_z\neq0$, the top and middle panels of Fig.~\ref{fig:intro}(e) show that the ansatz (\ref{eq:ansatz_eDQPT}) accurately captures the dynamics of the rate function, singular values, and overlaps. The avoided crossing of the singular values leads to a much faster growth of entanglement compared to pDQPTs and drives the switch of the transfer matrix eigenvalues: near the DQPT the quantum state undergoes a rearrangement whereby the initially off-diagonal component, which for $\lambda_2\ll \lambda_1$ provides a correction to the leading top-diagonal component, becomes the dominant contribution. Thus eDQPTs manifest a change in the leading component of the quantum state and can be revealed by the structure of non-local correlations, as we discuss below.

\begin{figure}[tb]
\begin{center}
\includegraphics[width=\columnwidth]{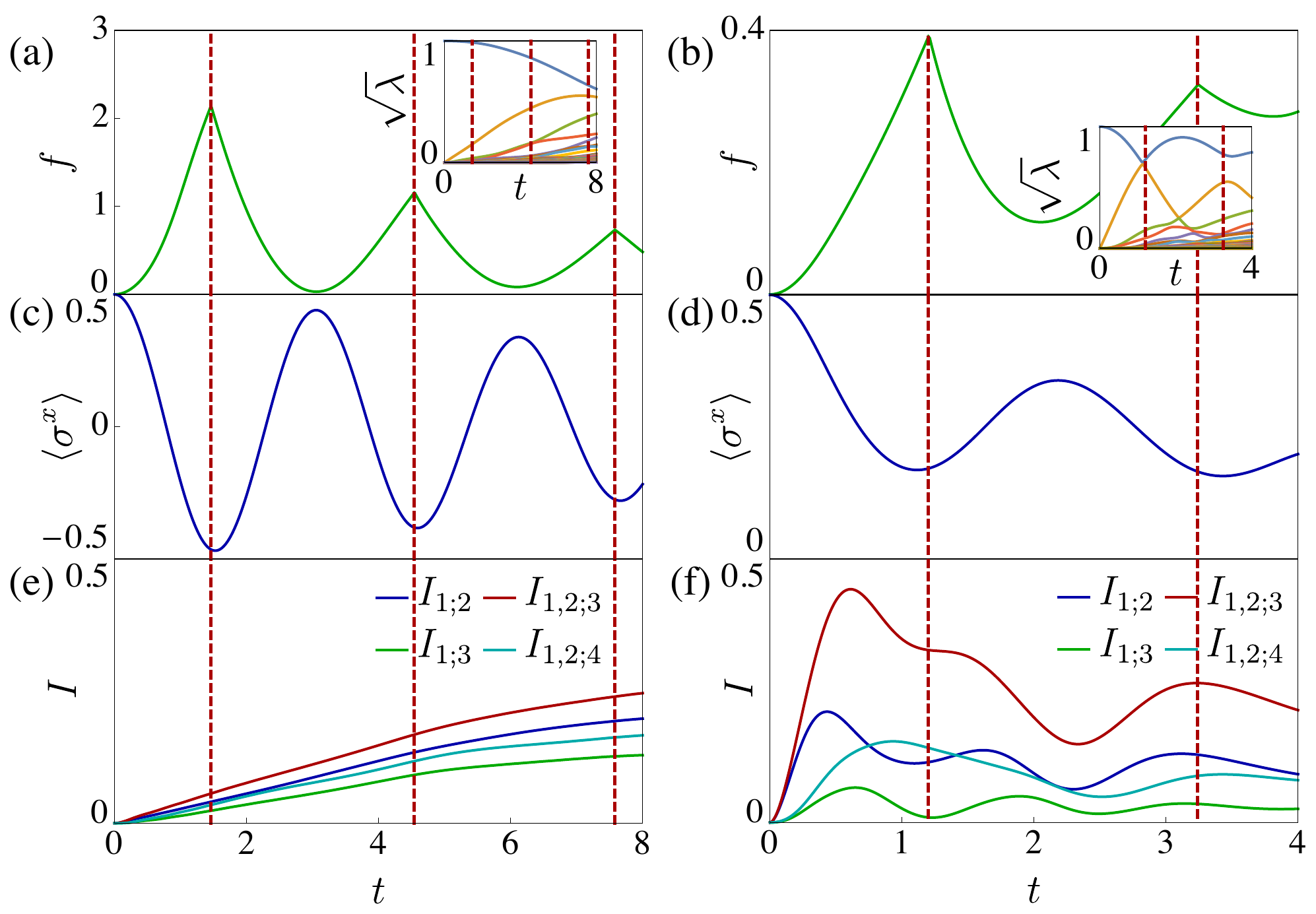}
\end{center}
\caption{\label{fig:observables} The qualitative behavior of the fidelity density is very similar for the pDQPTs in (a) and the eDQPTs in (b) that occur in the XXZ spin chain. In contrast, $x$-magnetization and MI have qualitatively different behavior for pDQPTs [panels (c) and (e)] and eDQPTs [panels (d) and (f)]. Simulations are performed using iTEBD with $\chi=200$. }
\end{figure}

\emph{DQPTs in the XXZ model.---}To demonstrate the existence of pDQPTs and eDQPTs  beyond the Ising chain, we consider quenches from the fixed initial product state $\ket{\psi_0}=\otimes_i \ket{\rightarrow}_i$. The dynamics is governed by the XXZ model with a field,
$H = \sum_{i,\alpha} \left[
J_\alpha \sigma_{i}^{\alpha}\sigma_{i+1}^{\alpha} +h_{\alpha}\sigma_{i}^{\alpha}
\right],
$
where $J_x=J_y$ and we set $h_y=0$. 
Figure~\ref{fig:observables}(a) shows dynamics for $J_x=J_y=0.9$, $J_z=1$, $h_x=0.1$, $h_z=1$, which displays pDQPTs, as it can be seen from the behavior of the entanglement spectrum in the inset. 
In Fig.~\ref{fig:observables}(b) we consider the same initial state evolved with $J_x=J_y=0.3$, $J_z=1$, $h_x=0.3$, $h_z=0.1$. In this case, cusps in $f(t)$ are close to avoided crossings in the entanglement spectrum (see inset), suggesting eDQPTs. The behavior of the overlaps shown in~\cite{SOM} confirms these expectations.

\emph{Experimental signatures.---}pDQPTs and eDQPTs have very different physical mechanisms, yet the fidelity density behaves qualitatively similarly, cf.~Fig.~\ref{fig:intro}(d)-(e) or Fig.~\ref{fig:observables}(a)-(b). An immediate distinction between different DQPTs is provided by the bipartite entanglement entropy $S$: pDQPTs are of semiclassical nature and occur in low-entanglement regions, whereas eDQPTs are triggered by avoided crossings in $\lambda_i$ at early times, reflected in rapid entanglement growth.

While local expectation values evolve smoothly and cannot indicate the precise location of DQPTs, they provide an additional test for the underlying physical mechanisms.
Namely, near pDQPTs the dominant component of the state is maximally far away from the initial state, thus the local magnetization along the orientation of the initial state has opposite sign compared to its value at $t=0$, see Fig.~\ref{fig:observables}(c). Near eDQPTs, which are characterized by larger entanglement, local expectation values are expected to be small; this is indeed confirmed by Fig.~\ref{fig:observables}(d), where the magnetization along the $x$-direction assumes its minimal magnitude near an eDQPT. 

The mutual information (MI) can be used to reveal the non-trivial entanglement pattern near eDQPTs. The MI between two regions $A$ and $B$ is defined as $I_{A;B} =S(A)+S(B)- S(A \cup B) $, where $S(\cdot)$ is the von Neumann entropy of a given region. Regions $A$, $B$ are chosen to contain one or two spins. Due to translational invariance only relative distances between regions is important. MI provides a basis-independent upper bound on connected correlation functions, which could reveal qualitatively similar behavior provided an appropriate basis is chosen.
Figure~\ref{fig:observables}(e) shows that the MI for all choices of regions $A$ and $B$ undergoes slow monotonic growth in the case of a pDQPT. In contrast, eDQPTs correspond to complex oscillatory dynamics of the MI; this is demonstrated in Fig.~\ref{fig:observables}(f), where DQPTs correspond to broad maxima in the MI $I_{1,2;3}$ between spins $\{1,2\}$ and $\{3\}$.

In~\cite{SOM} we show a similar pattern for the DQPTs of Fig.~\ref{fig:intro}(d)-(e) in the Ising model. The quick growth and non-monotonic behavior of the MI for eDQPTs signal a change in the dominant component of the wave function. The MI can be probed by  the connected correlation functions between the two subsystems, which are typically accessible in experiments.

\emph{Discussion.---} We introduced the notions of precession and entanglement DQPTs as two limiting cases, which have different underlying mechanisms and are associated to different physics. pDQPTs can be understood analytically by relying on the large entanglement gap $\lambda_1\gg \lambda_2$ and the dynamics being driven by single-spin terms in the Hamiltonian. In contrast, eDQPTs happen near avoided level crossings in the entanglement spectrum $\lambda_1\sim \lambda_2\gg\lambda_3$ and can also be analytically described by ignoring $\lambda_i$ with $i\geq 3$.

We demonstrated that pDQPTs and eDQPTs exist in different models. These two limits illustrate different physical mechanisms that cause DQPTs, whose relative importance can be qualitatively assessed from the behavior of local observables. Approximations with small bond dimension can then capture DQPTs, provided they incorporate the relevant physics; see \cite{SOM}. However, more complicated dynamics emerges when both precession and entanglement production are significant; for instance, in \cite{SOM} we deform eDQPTs into pDQPTs and show complicated hybrid behavior at intermediate couplings. 
This suggests that DQPTs are generically the outcome of a combination of factors; the interplay of different mechanisms may then be at the root of the rich phenomenology reported in the literature~\cite{heyl2018}.
Our work shows that focusing on the underlying mechanisms is a fruitful path to understanding DQPTs.
It would then be interesting to develop analytical ans\"atze to characterize situations with more than one dominant mechanism, as well as long-range interacting models~\cite{jurcevic2017,Halimeh2017,Homrighausen2017,zunkovic2018} and other cases that violate typical phenomenology~\cite{Homrighausen2017,Trapin2020,heyl2018}.

The connection between DQPTs and the spectrum of the fidelity transfer matrix, which is generically non-Hermitian, calls for exploring the relation between DQPTs and the theory of non-Hermitian matrices~\cite{Berg19,Ashida2020} that may allow a classification of DQPTs. Tensor network description could also be used to establish a notion of p- and eDQPTs in higher dimensions and understanding implications for string observables~\cite{halimeh2020local,bandyopadhyay2020observing}.

\emph{Acknowledgments.---}
SDN acknowledges funding from the Institute of Science and Technology (IST) Austria, and from the European Union's Horizon 2020 research and innovation programme under the Marie Sk\l{}odowska-Curie grant agreement No.~754411.  A.M.~and M.S.~were supported by the European Research Council (ERC) under the European Union's Horizon 2020 research and innovation programme (grant agreement No.~850899).

\clearpage
\pagebreak
\onecolumngrid
\begin{center}
\textbf{\large Supplementary material for ``Entanglement view of dynamical quantum phase transitions'' }\\[5pt]
\begin{quote}
{\small 
In this supplement we present further details on the analytical iMPS ans\"atze for p- and eDQPTs discussed in the main text, as well as additional results for the Ising and XXZ models. We also demonstrate the deformation of eDQPTs into pDQPTs upon varying the parameters of the quench Hamiltonian and explore quenches where both precession and entanglement generation are significant, revealing the complex phenomenology which arises in the intermediate regime between the contrasting pDQPT and eDQPT limits.
}\\[20pt]
\end{quote}
\end{center}
\setcounter{figure}{0}
\setcounter{table}{0}
\setcounter{page}{1}
\setcounter{section}{0}
\makeatletter
\renewcommand{\thefigure}{S\arabic{figure}}
\renewcommand{\thepage}{S\arabic{page}}

\twocolumngrid

\section{Canonical Form of MPS}

The iMPS representation of a many-body state is encoded in a tensor $A_{ij}^\sigma$ carrying a physical index $\sigma=\uparrow,\downarrow$ and bond indices $i,j=1,\dots,\chi$. Such representation is non-unique due to \textit{gauge invariance}: for any invertible $\chi \times \chi$ matrix $G$, the matrix $ \tilde{A}^\sigma_{ij} = [G A^\sigma G^{-1}]_{ij}$ provides an equivalent representation of the state.
This gauge freedom may be fixed by imposing additional conditions on the matrix $A$. A particularly convenient gauge fixing is given by the canonical form, which requires the tensor $A$ to be represented as $ A^\sigma_{ij} = \Lambda_{i} \Gamma_{ij}^\sigma$~\cite{Vidal2006,Orus2008}: the diagonal matrix $\Lambda_{ii}$ contains the singular values $\{ \sqrt{\lambda}_i \}$ of the Schmidt decomposition across a bond, whose squares yield the entanglement spectrum, while the tensor $\Gamma^\sigma_{ij}$ carries a physical index, so that its elements can be seen as (not necessarily normalized) spinors. The canonical form tensors $\Lambda$ and $\Gamma$ satisfy a set of constraints given in the main text.

\begin{figure}[tb]
\begin{center}
\includegraphics[width=\columnwidth]{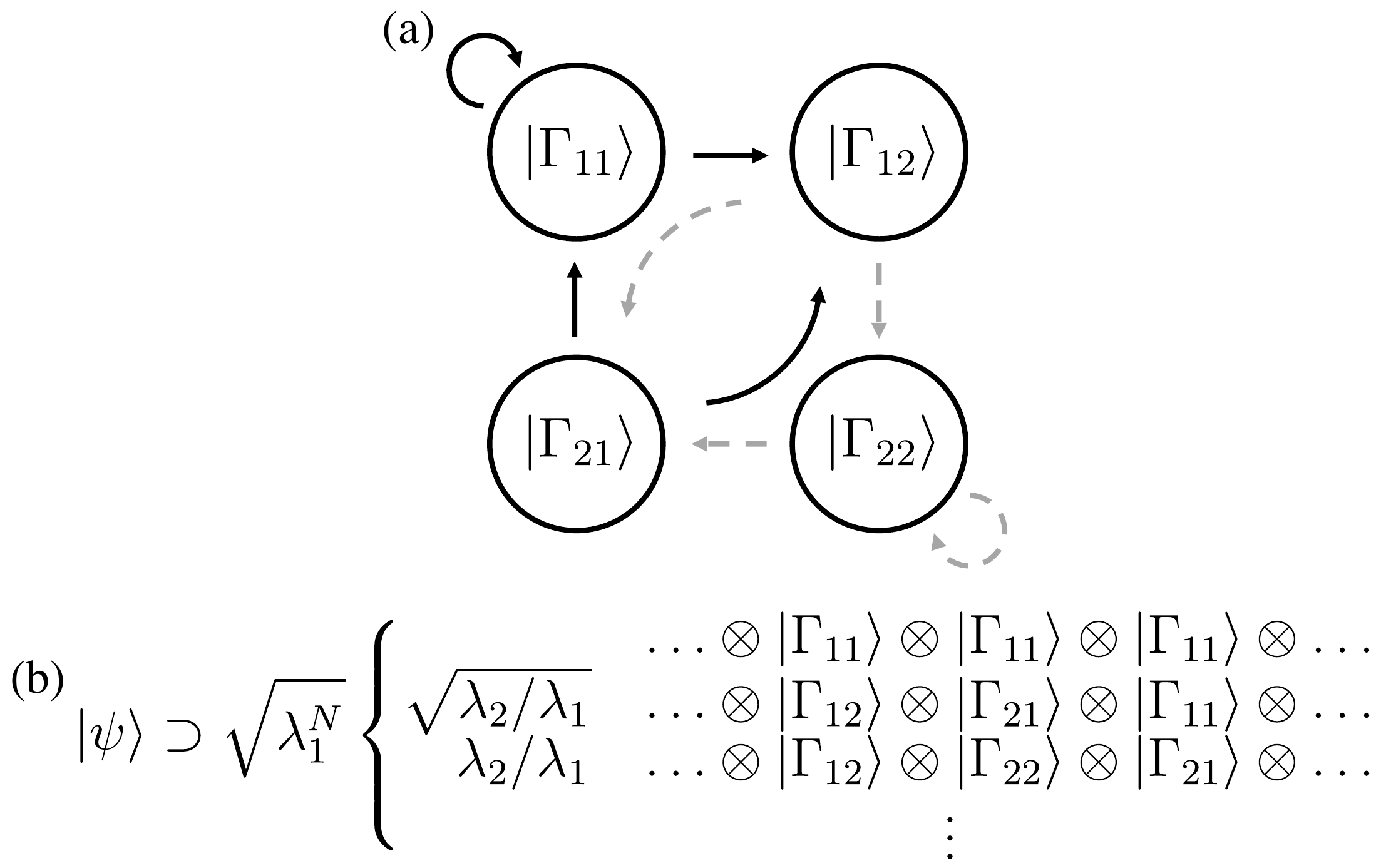}
\end{center}
\caption{(a) Automaton representation of an iMPS with $\chi=2$ written in the canonical form. Dark solid arrows carry a weight $\sqrt{\lambda_1}$ whereas light dashed arrows carry a weight $\sqrt{\lambda_2}\leq \sqrt{\lambda_1}$. (b) The iMPS can be seen as a linear superposition of all possible product states generated by the automaton.
\label{fig:automaton}}
\end{figure}

As discussed in the main text, the canonical form reveals the contributions of entanglement and precession, providing a tool to understand the driving physical mechanisms of DQPTs and how these are reflected in the behavior of other quantities. The structure encoded in a canonical form iMPS can be conveniently visualized by means of an \textit{automaton} picture~\cite{Crosswhite2008}, shown in Fig.~\ref{fig:automaton} for a $\chi=2$ iMPS state.  

Each circle in Fig.~\ref{fig:automaton}(a) corresponds to a site and carries a physical vector $\Gamma^\sigma_{ij}$, denoted as $\ket{\Gamma_{ij}}$ in the figure. The arrows give the allowed choices for the vector at the following site, weighted by the singular values $\sqrt{\lambda}_i$. In the case when $\lambda_1\gg \lambda_2$ that will be relevant for pDQPTs below, the dominant contribution is given by the $\ket{\Gamma_{11}}$ state, as each inclusion of $\ket{\Gamma_{12}}$ and other components is suppressed by at least a factor $\sqrt{\lambda_{2}/\lambda_1}\ll 1$. Thus in this limit the cartoon picture of the quantum state can be visualized as a dilute set of inclusions of $\ket{\Gamma_{12}}$, $\ket{\Gamma_{21}}$, and $\ket{\Gamma_{22}}$ into the dominant product state $ \otimes_i \ket{\Gamma_{11}}$. (Due to the orthogonality catastrophe, the overlap of the MPS state with the $ \otimes_i \ket{\Gamma_{11}}$ state vanishes in thermodynamic limit provided $\lambda_1<1$.)

In contrast, if one has $\lambda_1\approx \lambda_2$ (recall that $\lambda_1>\lambda_2$ by assumption), which is the relevant case for eDQPTs, the automaton provides a very different picture of the quantum state. The closeness of two singular values entails that a large number of excitations on top of the product state $ \otimes_i \ket{\Gamma_{11}}_i$ can be created at a small cost. Moreover, the avoided crossing in $\lambda$'s implies that the formerly subleading component of the MPS state will become dominant after such avoided crossing. Thus, at the points of avoided crossings a broad rearrangement of the quantum state is happening, whereby formerly subleading components become dominant.

\section{Analytical $\mbox{pDQPT}$ ansatz}\label{sec:pDQPTansatz}
\subsection{Effective Hamiltonian}
To illustrate pDQPTs, we construct an analytical ansatz capable of capturing the relevant physics. We consider quenches in the Ising model; the time-evolved state in the Schr\"odinger picture is then
\begin{align}
\ket{\psi(t)} = e^{-it H}\ket{\psi_0}
\end{align}
with $H$ given by~(\ref{eq:Hising}). Since pDQPTs are dominated by single-spin terms, it is convenient to split the Hamiltonian between the free-precessing part, $H_0= \sum_i[h_{x}\sigma_{i}^{x}+h_{z}\sigma_{i}^{z}]$, and the interaction, $V=J\sum_i \sigma_{i}^{z}\sigma_{i+1}^{z}$.
To account for the leading role of precession, we rewrite the dynamics in the rotating frame with respect to $H_0$. 
In the rotating frame, operators evolve according to
\begin{align}
\tilde{\mathcal{O}}(t) = e^{iH_0 t} \mathcal{O} e^{-it H_0}
\end{align}
where $\mathcal{O}$ is the (time-independent) Schr\"odinger picture operator.
Enforcing that expectation values be invariant upon switching to the rotating frame, $\bra{\tilde{\psi}(t)} \tilde{\mathcal{O}}(t) \ket{\tilde{\psi}(t)} = \bra{\psi(t)} \mathcal{O} \ket{\psi(t)}$, defines the rotating-frame state,
\begin{align}\label{eq:rotatingFrameState}
\ket{\tilde{\psi}(t)}=e^{iH_0 t} \ket{\psi(t)} ,
\end{align}
in terms of the Sch\"odinger picture state $\ket{\psi(t)}$.
The equation of motion satisfied by $\ket{\tilde{\psi}(t)}$ is readily obtained by differentiating~(\ref{eq:rotatingFrameState}):
\begin{align}\label{eq:rotatingFrameStateEvo}
\frac{d}{dt }\ket{\tilde\psi(t)}  = -i\tilde{V}\ket{\tilde\psi(t)} ,
\end{align}
where
\begin{align}\label{eq:sigmaTilde}
\tilde{V} = J \sum_i \tilde{\sigma}^z_i \tilde{\sigma}^z_{i+1}, \quad \tilde{\sigma^z} =  e^{i t H_0} \sigma^z e^{-i t H_0} \equiv \sum_a s_a \sigma^a .
\end{align}
The coefficients $s_x(t), s_y(t), s_z(t) $ are provided in the main text. One then has
$\ket{\tilde{\psi}(t)}= U_{\tilde{V}}(t) \ket{\psi_0}$, where $U_{\tilde{V}}(t) = \mathrm{T} e^{-i \int_0^t \tilde{V}(t^\prime)\dd t^\prime}$, so that the time-evolved state in the Schr\"odinger picture can be rewritten as
\begin{align}\label{eq:rotatingFrameEvo}
\ket{\psi(t)}  = e^{-i  H_0 t} U_{\tilde{V}} \ket{\psi_0} .
\end{align}
As it stands, Eq.~(\ref{eq:rotatingFrameEvo}) is an exact reformulation of the dynamics and is not amenable to direct evaluation. In order to obtain a closed-form approximation, we restrict our attention on quenches from the $\ket{\downarrow}$ product state, which is the ferromagnetic ground state for $h_x=0, h_z>0, J<0$. We consider the precession-dominated regime $h_x \gg h_z,J$, relevant for pDQPTs. In this regime, the entanglement growth along the $z$-axis is initially small, as demonstrated by comparing the connected correlations $C_{xx}$, $C_{yy}$, $C_{zz}$ defined as  $C_{ab} = \langle \sigma^a_i \sigma^b_{i+1} \rangle - \langle \sigma^a_i \rangle \langle \sigma^b_i \rangle$. Furthermore, terms featuring $\sigma^y$ are dominant with respect to those featuring $\sigma^x$ due to $|s_y|>|s_x|$. We take advantage of these observations to replace the operators $ \sigma^x$, $\sigma^z$ in~(\ref{eq:sigmaTilde}) by their expectation values, which in the present regime are well-approximated by free precession:
\begin{align}
\langle \sigma^{x,z}(t)\rangle \approx \bra{\downarrow} e^{iH_0 t} \sigma^{x,z} e^{-iH_0 t} \ket{\downarrow} = -s_{x,z}(t).
\end{align}
This approximation yields $\tilde{V}(t) \approx \tilde{V}_{\text{eff}}(t) = \sum_i [  J_{\text{eff}}(t) \sigma_i^y \sigma_{i+1}^y + h_{\text{eff}}(t) \sigma_i^y]$ up to an unimportant constant term, with
\begin{align}
J_{\text{eff}}&=J  s_y^2,\\
h_{\text{eff}}&= -2 Js_y (s_x^2+ s_z^2 ).
\end{align}
The approximate form of the operator $\tilde{V}(t)$ obtained above is still time-dependent, but is now made up of commuting terms, such that 
\begin{align}
\mathrm{T}e^{-i \int_0^t \tilde{V}(t^\prime) \dd t^\prime} \approx e^{-i \int_0^t \tilde{V}_{\text{eff}}(t^\prime) \dd t^\prime} = e^{-i t H_{\text{eff}}(t)}
\end{align}
where $t H_{\text{eff}}(t)  = J a(t)  \sum_i  \sigma^y_i \sigma^y_{i+1} + J b(t) \sum_i \sigma^y_i $, and $J a(t)=\int^t_0 J_{\text{eff}}(t^\prime) \dd t^\prime$, $J b(t)= \int_0^t h_{\text{eff}}(t^\prime)\dd t^\prime $ are explicitly given in the main text. 
At each time $t$, the time-evolved state can thus be equivalently obtained from an effective classical Hamiltonian $H_{\text{eff}}(t)$.

\subsection{Exponentiation of the effective Hamiltonian}
The effective Hamiltonian $H_{\text{eff}}$ is made up of commuting terms, so that its matrix exponential can be trivially factorized as a product over sites,
\begin{align}
e^{-itH_{\text{eff}}}=\prod_i e^{-i J a(t) \sigma^y_i \sigma^y_{i+1} - i J b(t) [\sigma^y_i +\sigma^y_{i+1} ]/2}.
\end{align}
One then inserts resolutions of the identity over pairs of neighboring sites, $\mathbbm
{1}_{i,i+1} = \mathbbm{1}_i \otimes \mathbbm{1}_{i+1}$ with $\mathbbm{1}_i= P^y_i + P^{-y}_i $, where we introduced the projectors on the $y$-eigenstates, $P_i^{\pm y }\equiv \ket{\pm y}_i \bra{\pm y}_i$, $\sigma^y \ket{\pm y} = \pm \ket{\pm y}$. This yields
\begin{align}\label{eq:resIdentity}
\begin{split} e^{-itH_{\text{eff}}}= \prod_i 
 \Big( & e^{-iJ  [a(t)+ b(t)]} P^y_i P^{y}_{i+1} +  e^{iJ a(t)} P^y_i P^{-y}_{i+1}  \\ 
& \, +e^{iJ a(t)} P^{-y}_i P^{y}_{i+1}+e^{-iJ  [a(t)- b(t)]}  P^{-y}_i P^{-y}_{i+1}  \Big) 
\end{split}
\end{align}
The sum of terms resulting from (\ref{eq:resIdentity}) can be reproduced by the $\chi=2$ MPO $e^{-itH_{\text{eff}}} = \prod_i U_i$ with
\begin{align}
U_i =\left(\begin{matrix}e^{-i J a(t) - i J b(t) }P^y_i & e^{i J a(t)  - i J b(t) }P^y_i  \\ e^{iJ a( t) + i J b(t)  } P^{-y}_i  & e^{-i J a(t) +  i J b(t) } P^{-y}_i \end{matrix} \right).
\end{align}
The exponentiation of the effective classical Hamiltonian $H_{\text{eff}}$ is reminiscent of the transfer matrix method used to solve classical Ising models~\cite{mussardo2009}, with the difference that the matrix elements in the present case  are operators rather than scalars.

Substituting the above in Eq.~(\ref{eq:rotatingFrameEvo}) and applying it to the $\downarrow$ initial state leads to the pDQPT MPS ansatz discussed in the main text. 
To bring this to canonical form, 
one computes the transfer matrix $T_{(ik)(jl)}=\sum_{\sigma} A^\sigma_{ij} A^{\sigma *}_{kl}$, where $(a,b)$ denotes a merging of  the $a,b$ virtual indices. From the transfer matrix we compute the left- and right- dominant eigenvectors $V_L$, $V_R$ that correspond to the eigenvalue with largest magnitude. These eigenvectors are reshaped as $\chi \times \chi $ matrices and decomposed as $V_L = Y^\dag Y $, $V_R = X X^\dag$~\cite{Orus2008}. Finally, the matrix $\Lambda$ is obtained from the singular value decomposition of the matrix $Y^TX = U \Lambda V$. The explicit calculation yields $\Lambda=\text{diag}(|\cos( Ja(t) )|,|\sin( Ja(t) )|)$. 
The tensor $\Gamma$ given in Eq.~(\ref{eq:Gamma-p}) is then obtained from $\Gamma_{ij}^\sigma = \sum_{kl}[V X^{-1}]_{ik} A^\sigma_{kl} [(Y^T)^{-1} U]_{lj} $~\cite{Orus2008}.

\subsection{Evolution of the overlaps}
\begin{figure}[tb]
\begin{center}
\includegraphics[width=0.7\columnwidth]{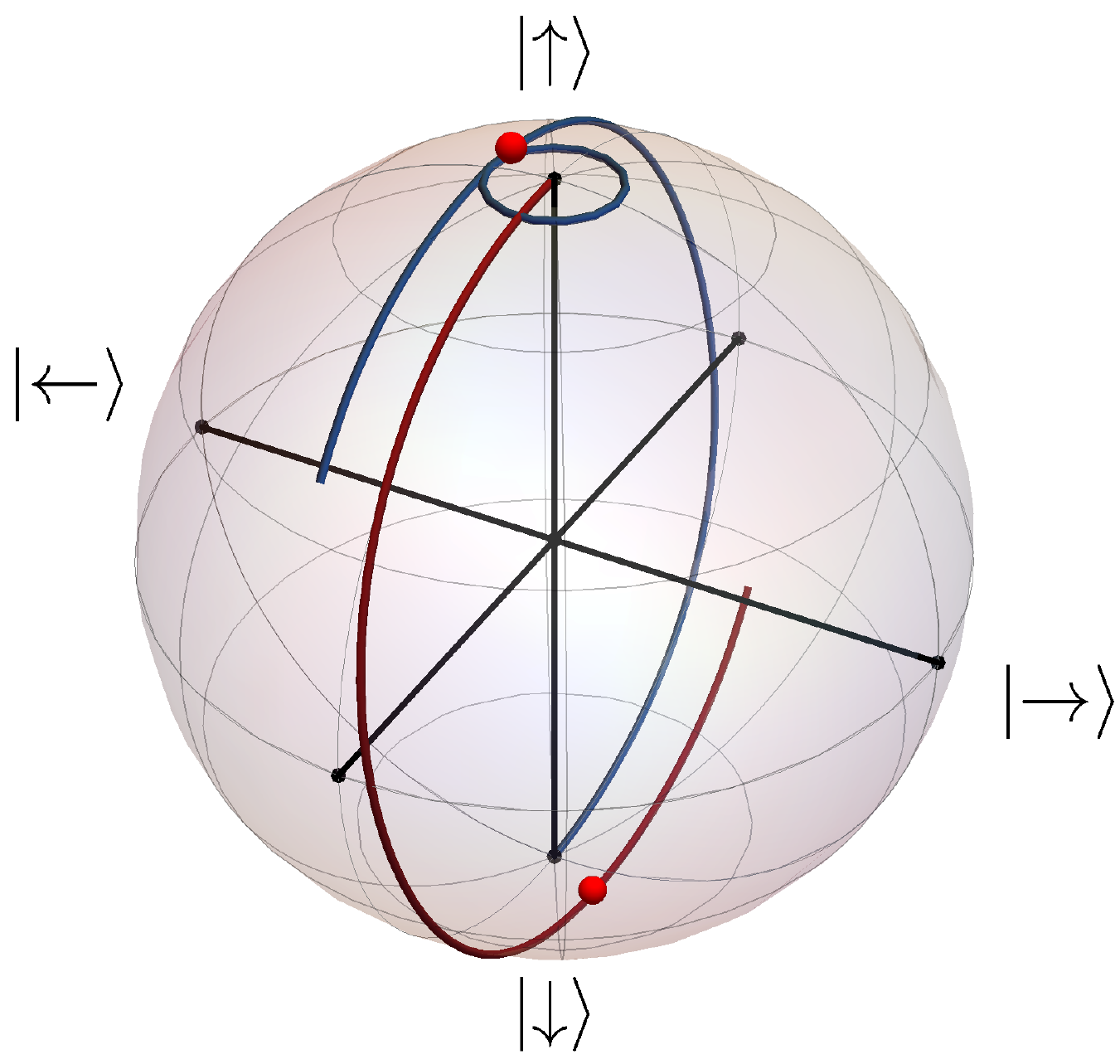}
\end{center}
\caption{Evolution of the vectors $\Gamma^\sigma_{11}$ (blue line), $\Gamma^\sigma_{12}=e^{i \phi}\Gamma^\sigma_{21}$ (red line), obtained from the analytical pDQPT ansatz~(\ref{eq:Gamma-p}), on the Bloch sphere. Since $\lambda_1 \gg \lambda_2$, $\Gamma^\sigma_{11}$ provides the dominant contribution to the MPS, with $\Gamma^\sigma_{12},\Gamma^\sigma_{21}$ giving subleading corrections (see Fig.~\ref{fig:automaton}). The vector $\Gamma^\sigma_{11}$ is initialized as $\ket{\downarrow}$, while $\Gamma^\sigma_{12}$ initially corresponds to $\ket{\uparrow}$. In the dynamics of Fig.~\ref{fig:intro}(b)-(d), these vectors perform simultaneous precession on the Bloch sphere. The precession of the dominant component away from the initial state gives rise to a pDQPT; the position of each vector on the respective trajectory at the time of the DQPT is marked by a red dot. In the vicinity of a pDQPT, $\Gamma^\sigma_{11}$ is nearly orthogonal to the initial state, so that $|o_{11}|$ approaches a minimum, while $\Gamma^\sigma_{12}$ is closest to the initial state and $|o_{12}|=|o_{\text{od}}|$ achieves a maximum. \label{fig:bloch}}
\end{figure}

In the present regime, one has $\lambda_2 \ll \lambda_1$ when the DQPT occurs (see Fig.~\ref{fig:intro}), so that the crossing of the transfer matrix eigenvalues is predominantly driven by the precession of the elements of $\Gamma$. The MPS ansatz shows that, to an excellent approximation, $\Gamma^\sigma_{ij}$ are normalized spinors; thus, the precession nature of pDQPTs can further understood by considering their evolution on the Bloch sphere, shown in Fig.~\ref{fig:bloch}. The dominant component $\Gamma_{11}^\sigma$, which at $t=0$ corresponds to the initial $\ket{\downarrow}$ product state, and the excitations $\Gamma^\sigma_{12}  = e^{i \phi} \Gamma^\sigma_{21}$, which initially correspond to the $\ket{\uparrow}$ state (orthogonal to the initial state), perform simultaneous precession, with DQPTs occurring when $\Gamma_{11}^\sigma \sim\ket{ \uparrow}$, $\Gamma_{12}^\sigma\sim \ket{\downarrow}$. 
The semiclassical nature of this driving mechanism and the role of excitations can be further understood by means of the automaton picture discussed in Fig.~\ref{fig:automaton}.

\section{Analytical $\mbox{eDQPT}$ ansatz}
\subsection{MPO form of the time evolution operator}
To construct an ansatz for eDQPTs, where interactions are expected to play a dominant role, we approximate the time-evolution operator by the second-order Trotter slicing $U(t) \approx U_L(t/2) U_I(t) U_L(t/2) $ with
\begin{align}\label{eq:trotterSplitting}
U_L(t) \equiv \prod_{i}e^{-it(h_{x}\sigma^{x}_{i}+h_{z}\sigma^{z}_{i})},\quad U_I(t) \equiv \prod_{i} e^{-it J \sigma_{i}^{z}\sigma_{i+1}^{z}},
\end{align} 
where $U_L$ captures local rotations while $U_I$ describes interactions. The interaction term is diagonal in the $\sigma_z$ basis, so that it admits the exact $\chi=2$ MPO representation
\begin{align}\label{eq:zzMPO}
U_I= \prod_i \left(\begin{matrix}e^{-iJt}\ket{\uparrow}\bra{\uparrow} & e^{i Jt}\ket{\uparrow}\bra{\uparrow}\\ e^{iJt }\ket{\downarrow}\bra{\downarrow} & e^{-iJt}\ket{\downarrow}\bra{\downarrow} \end{matrix} \right), 
\end{align}
similarly to the case discussed for the pDQPT ansatz. 
The full state can be readily obtained by applying the rotations $U_L(t/2)$ to the local states in (\ref{eq:zzMPO}), leading to the MPS ansatz (\ref{eq:ansatz_eDQPT}).
The corresponding canonical form for general $J,h_x,h_z$ can then be analytically obtained as discussed for the pDQPT ansatz. The analytical form of the tensor $\Gamma$ reveals a complicated structure, reflecting the involved behavior of the overlaps observed in Fig.~\ref{fig:intro}(e), while the entanglement spectrum $\{\lambda_i\}$, obtained by
squaring the diagonal of $\Lambda$, takes the relatively simple form provided in the main text.

\subsection{Canonical form of the MPS ansatz for the classical Ising model}
The tensor $\Gamma$ for the eDQPT ansatz can be greatly simplified in the limit $h_x=0$, which yields
\begin{align}
\Gamma= e^{-i t h_z \sigma^z} \left(\begin{matrix} \ket{\rightarrow} & i \ket{\leftarrow} \\
- \ket{\leftarrow} & -i\ket{\rightarrow} \end{matrix} \right)  ,
\end{align}
where the rotation  $e^{-it h_z \sigma^z}$ is applied to each of the spinor states in the matrix.
In this limit, the singular values reduce to $\sqrt{\lambda_{1,2}}= \{ |\cos(J t)|, |\sin (Jt)| \}$ and the MPS ansatz becomes an exact representation of the time-evolved state. 
Contraction with the complex-conjugated initial state $\ket{\rightarrow}$ gives the overlap matrix
\begin{align}\label{eq:overlapMatrix_classical}
o =\left(\begin{matrix} \cos (h_z t) &  \sin (h_z t) \\
i \sin (h_z t)&  -i \cos(h_zt) \end{matrix} \right) .
\end{align}

In spite of its apparent simplicity, this special case provides useful insights into e- and pDQPTs. 
In this limit, the dynamics of the state is fully factorized into the harmonic oscillations of the entanglement spectrum, with frequency $J$, and the free precession of the states in $\Gamma$, with frequency $h_z$. The eigenvalues of the fidelity transfer matrix are given by
\begin{align}\label{eq:TMevals_classical}
e_{1,2}= \frac{e^{-i J t}}{2}\left(  \cos (h_z t)\pm \sqrt{ e^{4 i J t}+\frac12\cos (2 h_z t)-\frac12}\right).
\end{align}
DQPTs arise whenever the difference in magnitude between $e_{1}$ and $e_2$ vanishes, $\Delta e =  |e_1| - |e_2|=0$. This corresponds to the condition
\begin{align}\label{eq:gapcondition}
\cos (h_z t)\, \text{Re} \left(\sqrt{2 e^{4 i J t} +\cos (2 h_z t)-1} \right) =0.
\end{align}
This condition is satisfied at the times $t=(n+1/2)\pi/(2J)$ when $\cos (2 h_z t)+2 e^{4 i J t}-1$ is real-valued and negative. 
It can be readily seen that these times corresponds to crossings in the entanglement spectrum, a hallmark of eDQPTs. 
However, Eq.~(\ref{eq:gapcondition}) is also satisfied whenever $\cos (h_z t)=0$. This condition corresponds to points of vanishing overlap $|o_{11}|$, given in~(\ref{eq:overlapMatrix_classical}), which are associated to pDQPTs. 
It can then be shown that there are no further solutions to Eq.~(\ref{eq:gapcondition}). Thus, in the present limit, DQPTs can be individually attributed to either precession or entanglement crossings, providing quintessential examples of p- and eDQPTs respectively.

This sheds light on the more general behavior observed when adding a finite $h_x$ such that the model is no longer solvable. In this case, the dynamics of overlaps and entanglement does not factorize into independent oscillations and depends on the values of all couplings. 
However, while smooth evolution of entanglement without particular signatures is still observed for pDQPTs, the entanglement avoided crossings which drive eDQPTs also affect the overlaps, inducing a complex behavior associated with a global rearrangement of the quantum state; see Fig.~\ref{fig:automaton} and Fig.~\ref{fig:deformation} below for further details. 

\begin{figure}[tb]
\begin{center}
\includegraphics[width=\columnwidth]{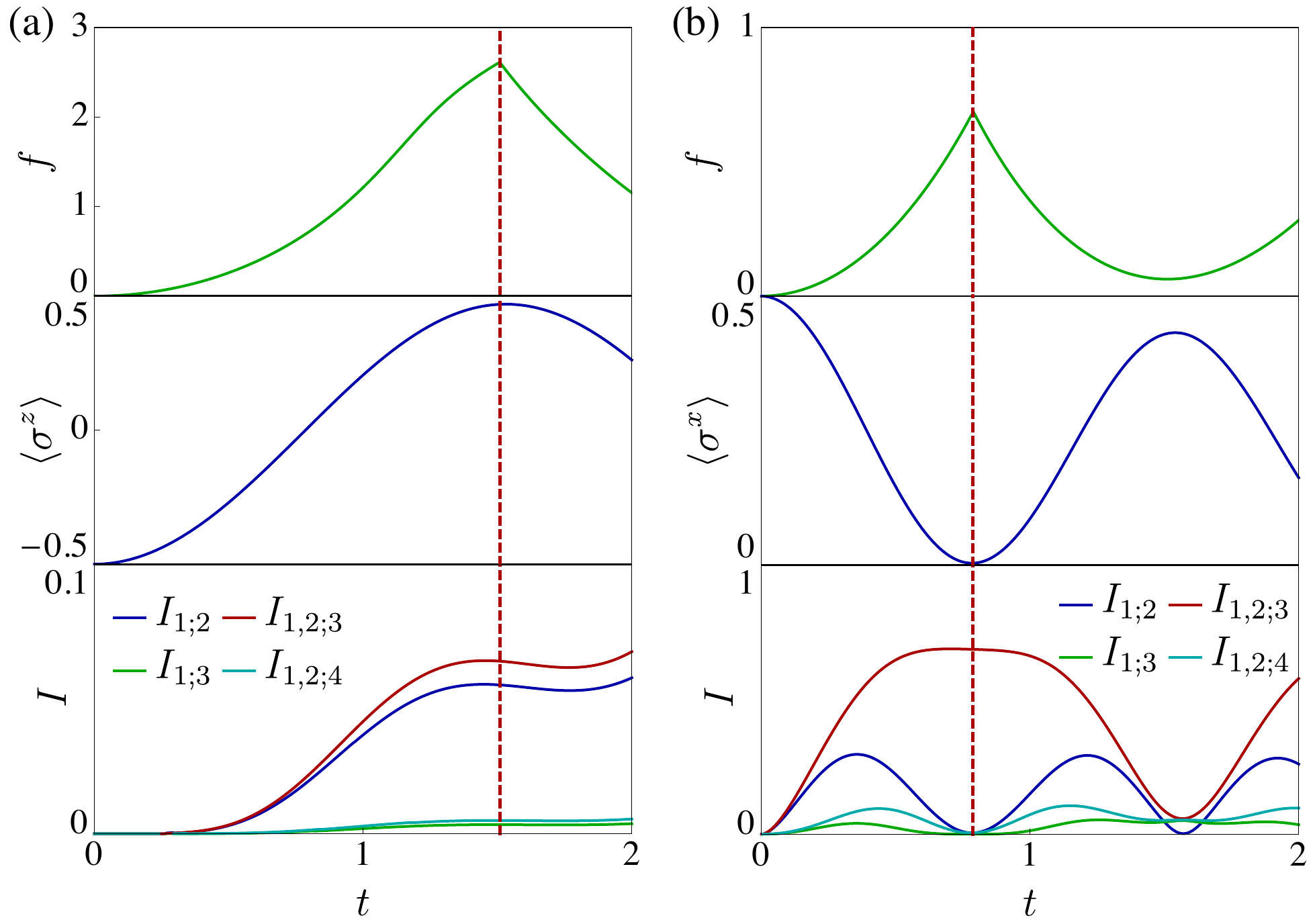}
\end{center}
\caption{\label{fig:extra_ising} Local magnetization in the direction of the initial state and mutual information for DQPTs in the quantum Ising model. DQPTs are marked by a dashed vertical line. Panel (a) shows a pDQPTs for the quench of Fig.~\ref{fig:intro}(b)-(d); in agreement with the discussion of the main text, this is accompanied by $\langle \sigma^z(t_\text{DQPT}) \rangle \approx - \langle \sigma^z(0)\rangle$, since the dominant vector $\Gamma^\sigma_{11}$ at this time has precessed maximally away from the initial state, while the mutual information shows a simple pattern of slow, approximately monotonic growth. In contrast, at the eDQPTs of panel (b), corresponding to the quench of Fig.~\ref{fig:intro}(c)-(e), the magnetization takes a value closest to zero, while the mutual information displays a complex pattern of correlations where two-body terms and $I_{1,2;4}$ reach a minimum while $I_{1,2;3}$ attains a plateau. Moreover, all MIs at the eDQPT are an order of magnitude larger compared to the pDQPT case. }
\end{figure}

\section{Contrasting and connecting $\mbox{eDQPTs}$ and $\mbox{pDQPTs}$}
\subsection{Experimental Signatures in the Ising Model}
The different nature of pDQPTs and eDQPTs was first illustrated in Fig.~\ref{fig:intro} by contrasting the behavior of entanglement and overlaps in the quantum Ising model. We then discussed how this difference is reflected in experimentally measurable quantities considering the XXZ model. In Fig.~\ref{fig:extra_ising} we additionally show that the same features can also be observed for the Ising model, considering the quenches of Fig.~\ref{fig:intro}; the behavior of the magnetization in the direction of the initial state and of the mutual information between different subsystems shows pronounced differences in the two cases, which can be understood in light of the proposed physical pictures.

\begin{figure}[tb]
\begin{center}
\includegraphics[width=\columnwidth]{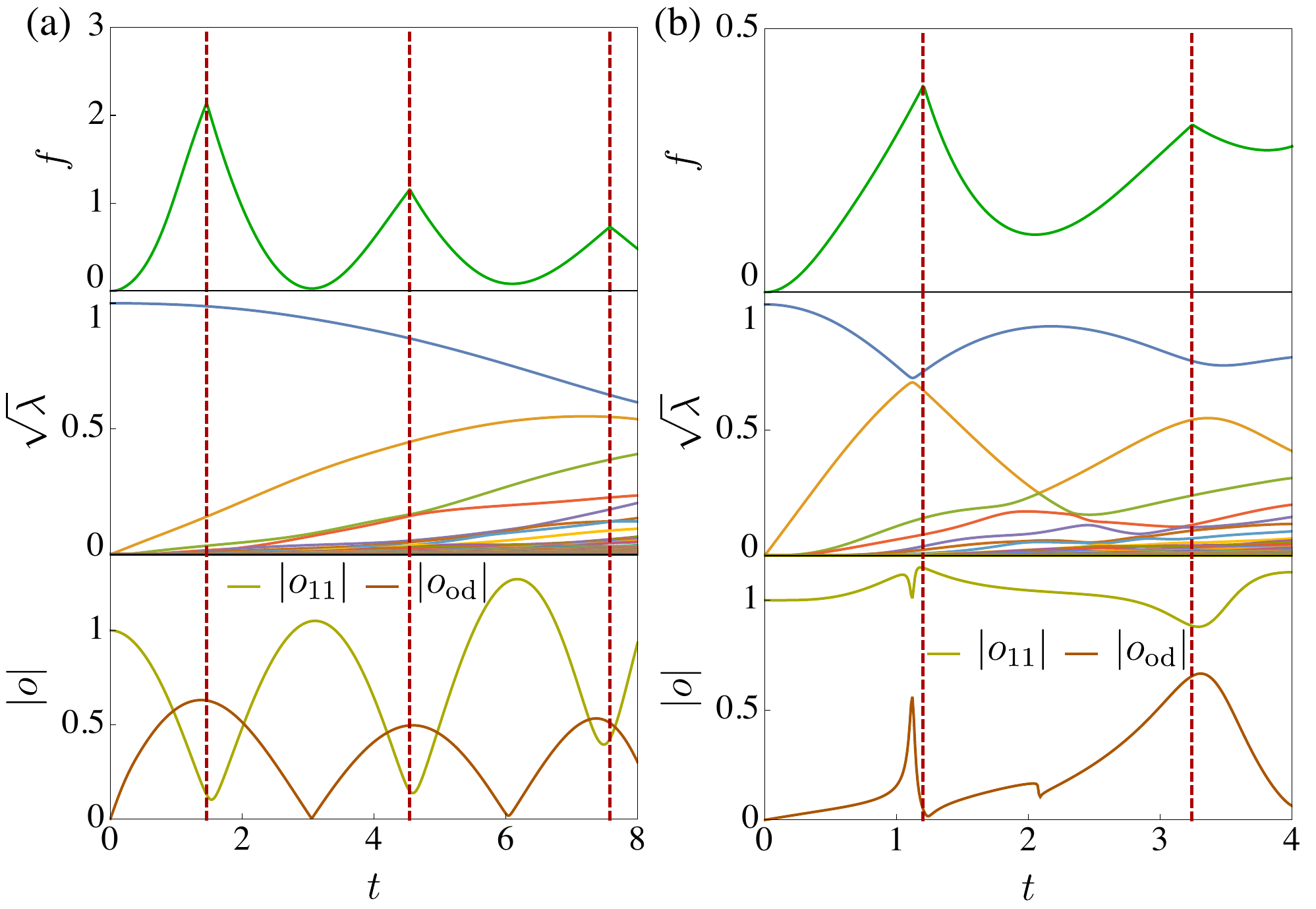}
\end{center}
\caption{\label{fig:extra_XXZ} Entanglement spectrum and overlaps for DQPTs in the XXZ model. Panel (a) corresponds to Fig.~\ref{fig:observables}(a) and shows pDQPTs; these are accompanied by a minimum of $|o_{11}|$, while the entanglement spectrum displays a gap and no particular signatures at DQPTs. Panel (b) shows the eDQPTs of Fig.~\ref{fig:observables}(b); in this case, DQPTs occur in the vicinity of an avoided crossing in the entanglement spectrum, with the overlaps also displaying avoided crossing behavior. }
\end{figure}

\begin{figure*}[t]
\begin{center}
\includegraphics[width=\columnwidth]{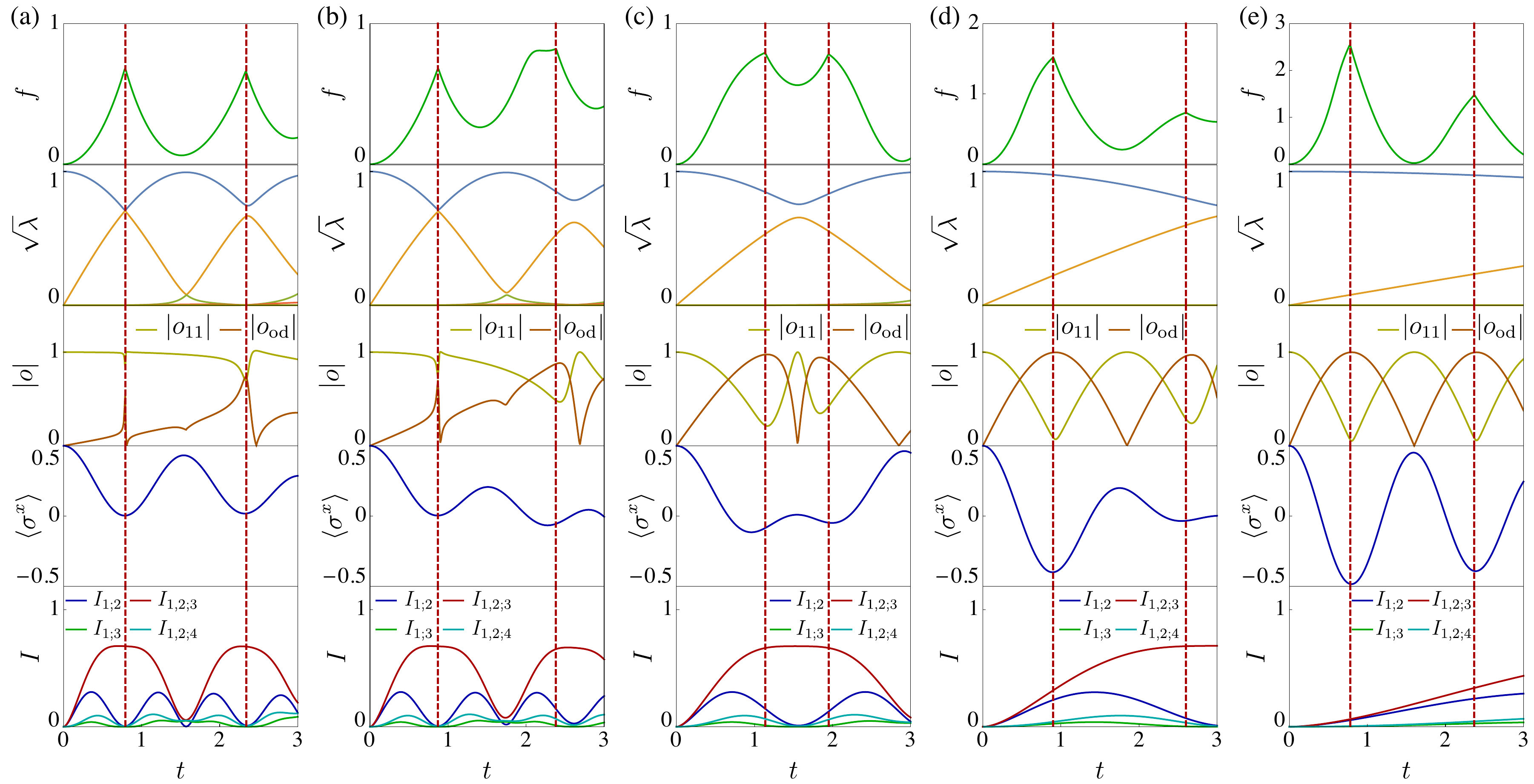}
\end{center}
\caption{\label{fig:deformation} Deformation of eDQPTs into pDQPTs. We choose the initial state $\ket{\rightarrow}$ and consider dynamics under the Ising Hamiltonian~(\ref{eq:Hising}), keeping $h_x$ = 0.1 constant and varying the values of $h_z$, $J$. We perform iTEBD truncating $\sqrt{\lambda_i}<10^{-9}$, which leads to a maximum bond dimension $\chi=12$ for the present quenches.\\
 (a) $h_z=0.15$, $J=1$; this is the same quench as Fig.~\ref{fig:intro}(c)-(e), but here we show the full entanglement spectrum and extend the time range so as to include a second DQPT. We observe two eDQPTs, characterized by entanglement avoided crossings and a complex behavior of the overlaps $|o_{11}|$, $|o_{\text{od}}|$, which approach each other at the DQPTs. This is also reflected in the magnetization $\langle \sigma^x \rangle$ approaching zero and the complex pattern displayed by the mutual information.\\
 (b) $h_z=0.35$, $J = 0.9$; as $h_z$ is increased and $J$ is reduced, the second entanglement avoided crossing is widened and the behavior of the second DQPT shifts away from the avoided crossing in entanglement spectrum. \\
 (c) $h_z=1.15$, $J = 0.5$; precession plays now an important role, and DQPTs manifest features of both e- and pDQPTs: they are associated with minima in $|o_{11}|$, which are however far from zero, and occur to the two sides of an entanglement avoided crossing, with observables suggesting a predominance of eDQPT nature.\\
 (d) $h_z = 1.65$, $J= 0.25$; as the field begins to dominate the dynamics, the entanglement gap widens and DQPTs start to acquire a pDQPT character, occurring near the minima of $|o_{11}|$.\\
 (e) $h_z = 1.95$, $J = 0.1$; in the strong-field regime, one retrieves two clean examples of pDQPTs: these are revealed by deep minima in $|o_{11}|$ and the large gap in the entanglement spectrum. Furthermore, in agreement with the general features of pDQPTs, the magnetization at DQPTs is approximately opposite to its initial value and the mutual information shows slow, featureless growth.}
\end{figure*}

\subsection{Entanglement and Overlaps in the XXZ Model}
The patterns in the entanglement spectrum and overlaps corresponding to p- and eDQPTs  are not restricted to the Ising model. To illustrate this, in Fig.~\ref{fig:extra_XXZ} we show the entanglement spectrum and overlaps for the XXZ model, considering the quenches of Fig.~\ref{fig:observables}. The observed behavior is in agreement with the general discussion of the main text, with pDQPTs corresponding to minima in $|o_{11}|$ while eDQPTs are associated with entanglement avoided crossings.

\subsection{Deforming eDQPTs into pDQPTs}

The concepts of pDQPTs and eDQPTs introduced in the main text provide two limiting cases in which DQPTs can be clearly ascribed to different physical mechanisms. For generic DQPTs, the situation can however be more involved, as both precession and entanglement production mechanisms might play a significant role. Part of the complex phenomenology reported in the literature might thus originate from the interplay of the discussed classes of DQPTs.

To illustrate this, we show how eDQPTs can be deformed into pDQPTs as a function of the Hamiltonian parameters by considering the Ising model. In Fig.~\ref{fig:deformation}(a), we begin by considering the same quench as in Fig.~\ref{fig:intro}(c)-(e) using the initial state $\ket{\rightarrow}$. As the longitudinal field $h_z$ is increased at the expense of the interaction strength $J$, precession also becomes important and DQPTs shifts away from the avoided crossing in the entanglement spectrum, see panel~(b). After a complex intermediate regime where DQPTs show a hybrid behavior, panel~(c), they gradually acquire the characteristics of pDQPTs, as shown in panel~(d). Finally, pDQPT features become very pronounced in the strong-field regime, as demonstrated in panel~(e).

\subsection{Reversing the Quench Direction}

\begin{figure}[t]
\begin{minipage}[c]{0.48\textwidth}
\begin{center}
\includegraphics[width=\columnwidth]{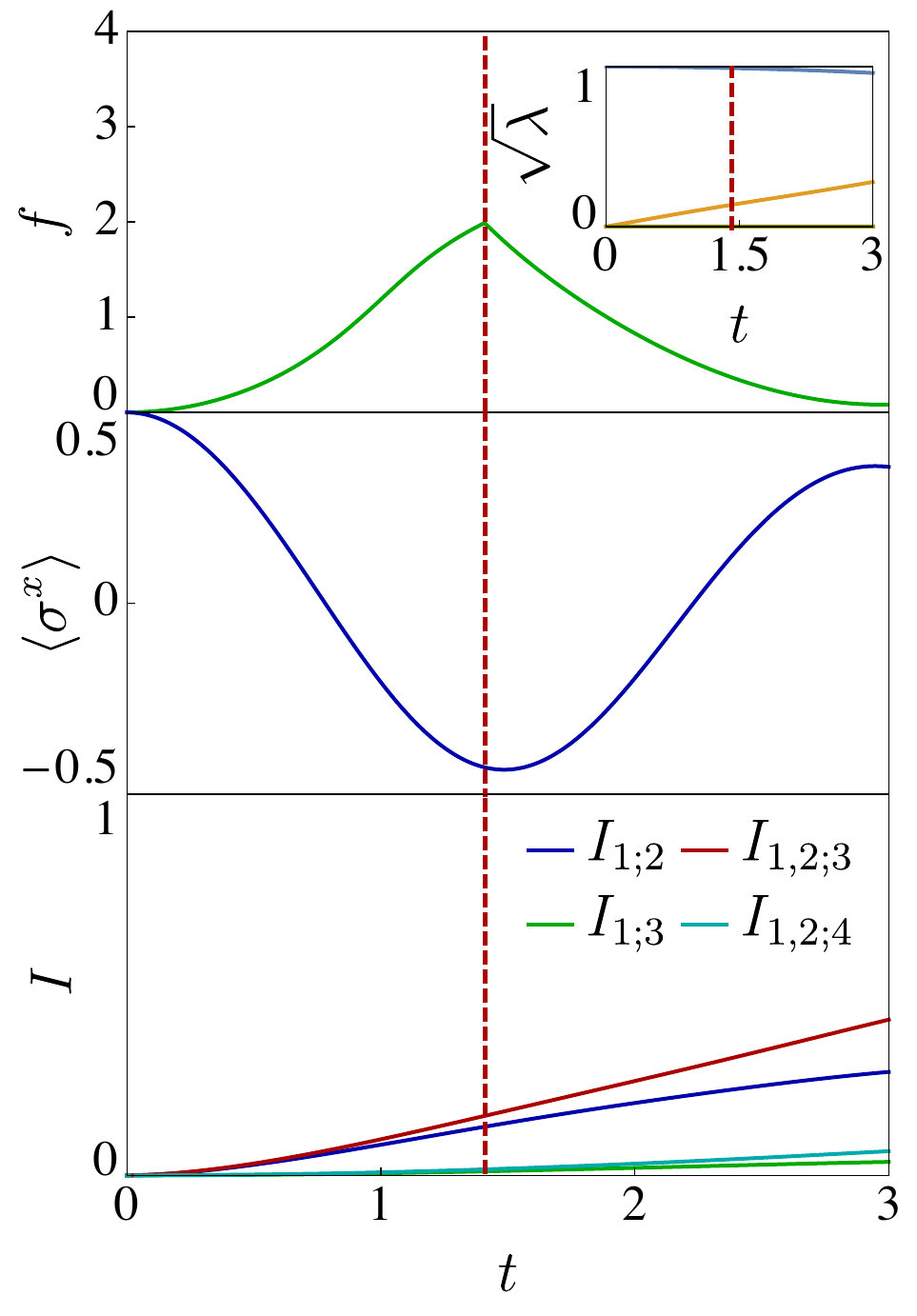}
\end{center}
\end{minipage}
\caption{The panels show, top to bottom: fidelity density $f$, $x$-magnetization $\langle \sigma^x \rangle$ and mutual information $I$ for the quench from the $\ket{\rightarrow}$ product state evolved with the Ising Hamiltonian with $J=0.1$, $h_x=0.15$, $h_z=1$. The inset further shows the entanglement spectrum. The observed phenomenology is consistent with a pDQPT.
\label{fig:reversed_quench} }
\end{figure}

Below we consider the sensitivity of the nature of DQPTs to a reversal in the quench direction. We use a pDQPT example for the Ising model discussed in the main text. The system is initialized in the $\ket{\downarrow}$ product state, corresponding to the ground state of the Ising Hamiltonian [Eq.~(\ref{eq:Hising}) in the main text] with $J<0$, $h_x=0$, $h_z>0$. The time evolution is performed using the Ising Hamiltonian with $J=0.1$, $h_x=1$, $h_z=0.15$.

The initial state in the above quench corresponds to the ferromagnetic phase of the Ising model with $Z_2$ symmetry being broken by the longitudinal field, while the Hamiltonian that governs the time evolution has a paramagnetic ground state. In order to reverse the quench direction, we start with a ground state of paramagnet-type, the $\ket{\rightarrow}$ product state. (Note that we checked that using the ground state of the Hamiltonian with finite but small values of $J$, $h_z$ does not not lead to qualitative differences.)

One option is to quench to the ferromagnetic phase with weak $Z_2$ symmetry breaking. This amounts to performing the time evolution with a classical Hamiltonian where $J$ is dominant, for instance $J=-1$, $h_x=0$, $h_z=0.1$. Such quench leads to eDPQTs, as we demonstrated in the main text and above. Another possibility is to quench into the ferromagnetic phase with strongly broken $Z_2$ symmetry, when $h_z \gg J$ is the dominant term in the Hamiltonian. Such quench results in pDQPT physics, see Fig.~\ref{fig:reversed_quench}.  

Thus, we conclude that reversing the quench direction can result in either eDQPT or pDQPT physics, depending on the relative strength of spin-spin interactions and single-spin terms in the Ising model; this is in agreement with the general picture discussed in the main text.

\begin{figure}[th!]
\begin{center}
\includegraphics[width=\columnwidth]{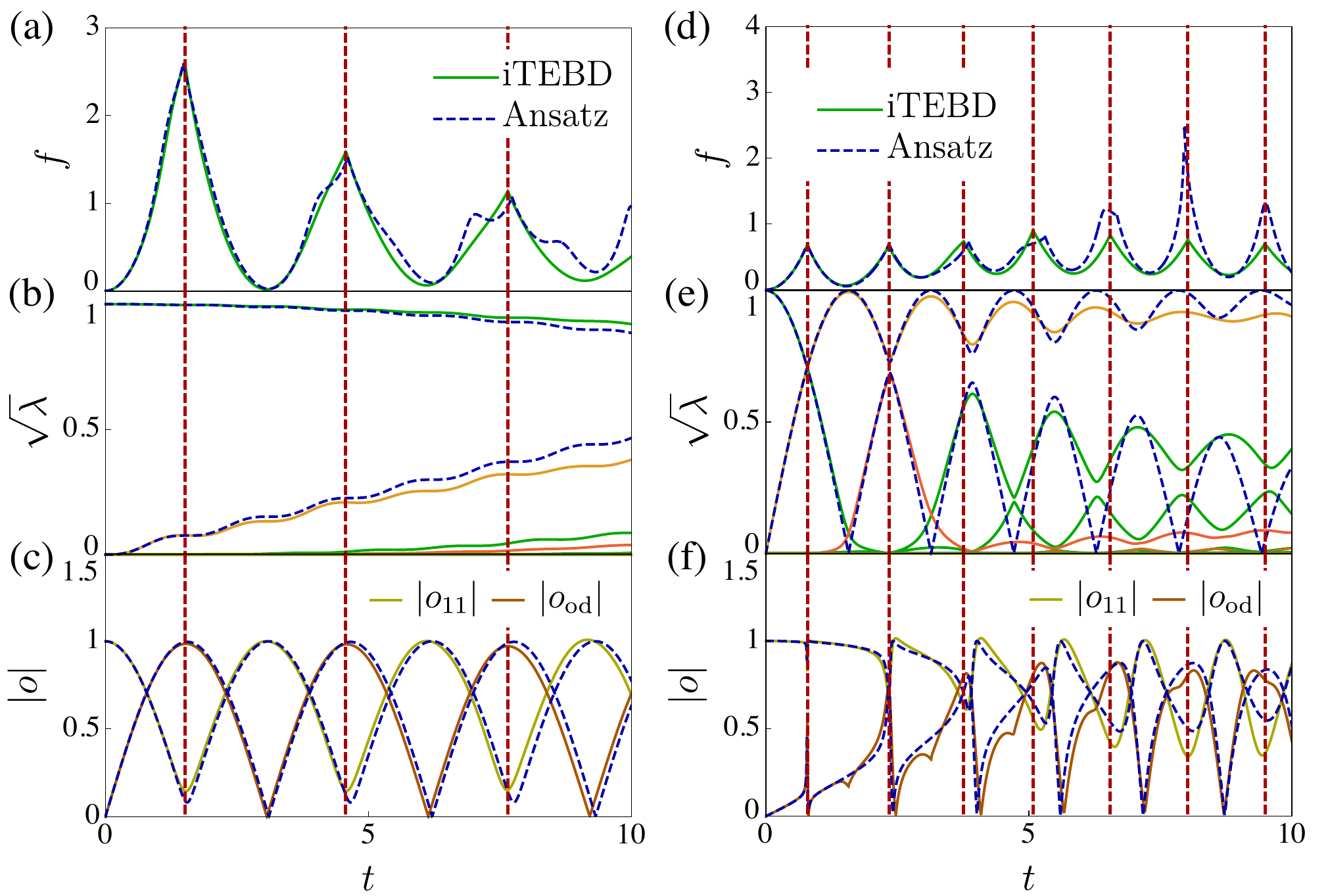}
\end{center}
\caption{Extended time evolution for the quenches of Fig.~\ref{fig:intro}(d) [panels (a)-(c)] and Fig.~\ref{fig:intro}(e) [panels (d)-(f)]. We perform iTEBD time evolution truncating $\sqrt{\lambda}<10^{-9}$, which leads to a maximal bond dimension $\chi=19$ and $\chi=28$ respectively. The analytical ans\"atze introduced in the main text (dashed blue lines), in spite of their much smaller bond dimension $\chi=2$, qualitatively capture the behavior of the DQPTs, predicting their occurrence and location with good accuracy. This shows that in the present cases, which provide prototypical examples of p- and eDQPTs respectively, it is sufficient to well-approximate the behavior of the dominant two components of the entanglement spectrum (b), (e) and the dominant and off-diagonal overlaps (c), (f) to capture the behavior of DQPTs.
\label{fig:long_time_ising} }
\end{figure}

\begin{figure}[th!]
\begin{center}
\includegraphics[width=\columnwidth]{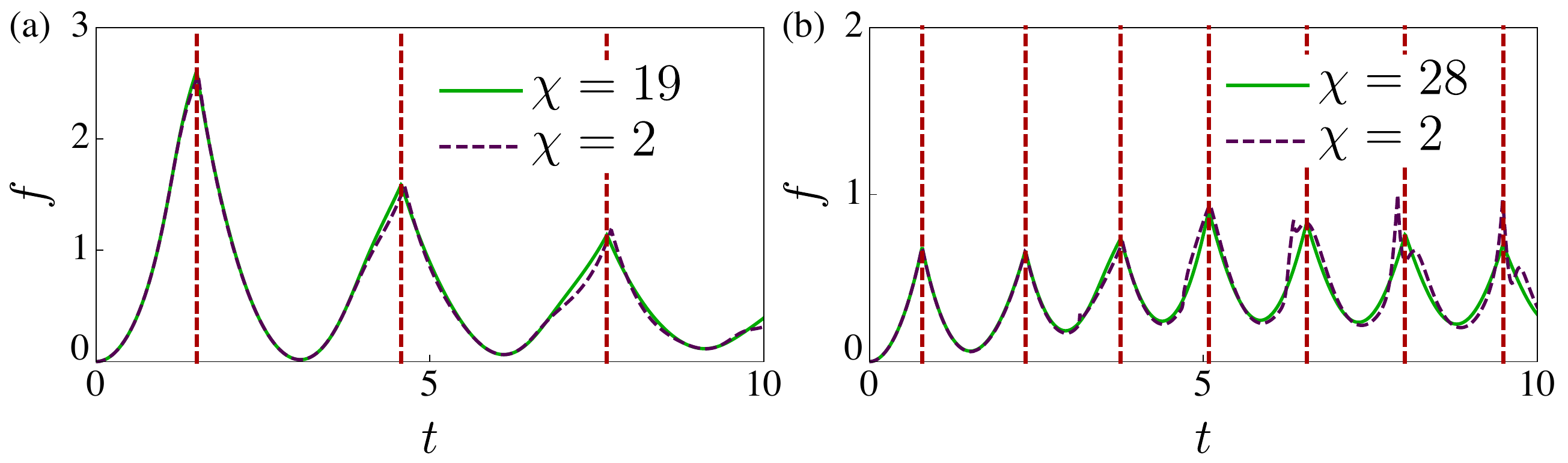}
\end{center}
\caption{\label{fig:long_time_ising_trunc} Comparison of the fidelity density obtained from the full time evolved state (full lines) and truncating the time-evolved state to a $\chi=2$ MPS (dashed lines) for the quenches shown in (a) Fig.~\ref{fig:intro}(d) and (b) Fig.~\ref{fig:intro}(e). The fidelity obtained from the truncated state is in good agreement with the numerically exact result and correctly captures the occurrence of DQPTs, approximately predicting their location. This shows that in the present cases, where either precession or entanglement production dominates, a $\chi= 2$ approximation of the state is capable of capturing DQPTs.}
\end{figure}

\section{DQPTs of strongly entangled states}
Considering the quantum Ising model, we have shown how $\chi=2$ analytical ans\"atze can be used to reveal the physics underlying DQPTs, leading to the definition of the limiting cases of p- and eDQPTs. 
The ans\"atze are capable to correctly capture DQPTs even though the numerics we benchmark them against have larger bond dimension. 
The question then naturally arises as to the range of applicability of $\chi=2$ approximations. We conjecture this is closely related to the existence of a dominant driving mechanism for DQPTs for a given quench. 

\begin{figure}[t]
\begin{center}
\includegraphics[width=\columnwidth]{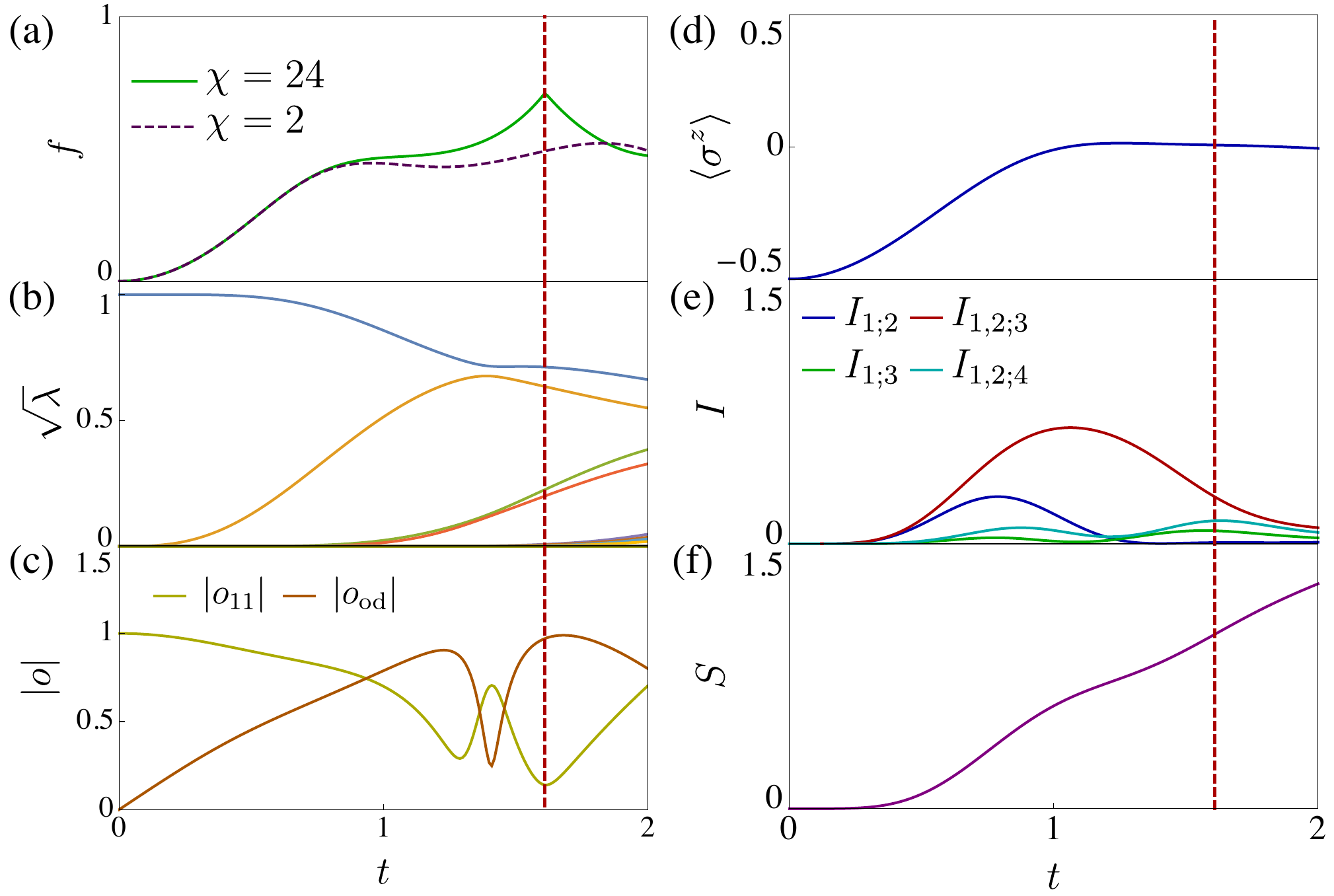}
\end{center}
\caption{Dynamics of the initial product state $\otimes_i \ket{\downarrow}$ evolved with the Ising Hamiltonian~(\ref{eq:Hising}) with couplings $J=h_x=h_z=1$. We perform iTEBD truncating $\sqrt{\lambda_i}<10^{-9}$, which results in a maximum bond dimension $\chi=24$ at $t=2$. In this regime, precession and entanglement production are both significant, so that pDQPT and eDQPT characters are blurred. (a) A DQPT occurs in the fidelity density (full line); however, in contrast with Fig.~\ref{fig:long_time_ising_trunc}, the approximate fidelity obtained by truncating the time-evolved state to $\chi=2$ (dashed line) fails to predict a DQPT, showing that for this quench it is necessary to retain $\lambda_i$ with $i>2$. The DQPT occurs (b) following an avoided crossing in the entanglement spectrum, and (c) at a minimum of $|o_{11}|$ and maximum of $|o_{\text{od}}|$; these findings illustrate that both p- and eDQPT driving mechanisms are simultaneously at play. In contrast to the pure pDQPT case, the overlaps show a complicated time-evolution. In contrast to the pure eDQPT case, the DQPT occurs when $|o_{11}|$ and $|o_{\text{od}}|$ are maximally different, rather than comparable. The behavior of both (d) the magnetization in the direction of the initial state and (e) the mutual information suggest a prevalence of eDQPT character. The entanglement entropy (f) shows rapid, nearly featureless growth.
\label{fig:ising_equal_couplings}  }
\end{figure}

In Fig.~\ref{fig:long_time_ising}, we show that the ans\"atze are capable to capture DQPTs for the quenches in Fig.~\ref{fig:intro}(d) and (e), which provide prototypical examples of p- and eDQPTs, also at later times. In spite of the numerically exact iTEBD dynamics having even larger bond dimension [$\chi=19$ for (a), (b), (c) and $\chi=28$ for (e), (f), (g)], the $\chi=2$ ans\"atze introduced in the main text still approximately predict the location of DQPTs. This can be attributed to the good approximation the ans\"atze yield to the dominant elements of the entanglement spectrum, (b) and (e), and the dominant overlaps, (c) and (f).
The fact that  for the quenches of Fig.~\ref{fig:intro}(d)-(e) DQPTs are predominantly determined by the top $2 \times 2$ component of the state can be further demonstrated by truncating the full iTEBD time-evolved state to $\chi= 2$, which is equivalent to discarding $\lambda_i$ for $i>2$ and renormalizing the state. Fig.~\ref{fig:long_time_ising_trunc} shows that this approximate state is sufficient to capture DQPTs, even though the full time-evolution requires significantly larger bond dimension.

Finally, in Fig.~\ref{fig:ising_equal_couplings} we consider a parameter range where both mechanisms are comparable, so that the distinction between pDQPTs and eDQPTs is blurred. In this case, a truncation of the MPS state obtained from iTEBD evolution with $\chi=24$ to $\chi=2$ leads to a disappearance of the DQPT.
In contrast, in Fig.~\ref{fig:long_time_ising}(b), where the eDQPT mechanism dominates, we saw a $\chi=2$ approximation correctly capturing DPQTs for a state with $\chi=28$.
Thus, we attribute the failure of the truncated $\chi=2$ state in capturing the DQPT to the intermediate character of the DQPT, which is driven by both precession and entanglement mechanisms.


\begin{thebibliography}{52}%
\makeatletter
\providecommand \@ifxundefined [1]{%
 \@ifx{#1\undefined}
}%
\providecommand \@ifnum [1]{%
 \ifnum #1\expandafter \@firstoftwo
 \else \expandafter \@secondoftwo
 \fi
}%
\providecommand \@ifx [1]{%
 \ifx #1\expandafter \@firstoftwo
 \else \expandafter \@secondoftwo
 \fi
}%
\providecommand \natexlab [1]{#1}%
\providecommand \enquote  [1]{``#1''}%
\providecommand \bibnamefont  [1]{#1}%
\providecommand \bibfnamefont [1]{#1}%
\providecommand \citenamefont [1]{#1}%
\providecommand \href@noop [0]{\@secondoftwo}%
\providecommand \href [0]{\begingroup \@sanitize@url \@href}%
\providecommand \@href[1]{\@@startlink{#1}\@@href}%
\providecommand \@@href[1]{\endgroup#1\@@endlink}%
\providecommand \@sanitize@url [0]{\catcode `\\12\catcode `\$12\catcode
  `\&12\catcode `\#12\catcode `\^12\catcode `\_12\catcode `\%12\relax}%
\providecommand \@@startlink[1]{}%
\providecommand \@@endlink[0]{}%
\providecommand \url  [0]{\begingroup\@sanitize@url \@url }%
\providecommand \@url [1]{\endgroup\@href {#1}{\urlprefix }}%
\providecommand \urlprefix  [0]{URL }%
\providecommand \Eprint [0]{\href }%
\providecommand \doibase [0]{https://doi.org/}%
\providecommand \selectlanguage [0]{\@gobble}%
\providecommand \bibinfo  [0]{\@secondoftwo}%
\providecommand \bibfield  [0]{\@secondoftwo}%
\providecommand \translation [1]{[#1]}%
\providecommand \BibitemOpen [0]{}%
\providecommand \bibitemStop [0]{}%
\providecommand \bibitemNoStop [0]{.\EOS\space}%
\providecommand \EOS [0]{\spacefactor3000\relax}%
\providecommand \BibitemShut  [1]{\csname bibitem#1\endcsname}%
\let\auto@bib@innerbib\@empty
\bibitem [{\citenamefont {Langen}\ \emph {et~al.}(2015)\citenamefont {Langen},
  \citenamefont {Geiger},\ and\ \citenamefont
  {Schmiedmayer}}]{reviewColdAtoms2015}%
  \BibitemOpen
  \bibfield  {author} {\bibinfo {author} {\bibfnamefont {T.}~\bibnamefont
  {Langen}}, \bibinfo {author} {\bibfnamefont {R.}~\bibnamefont {Geiger}},\
  and\ \bibinfo {author} {\bibfnamefont {J.}~\bibnamefont {Schmiedmayer}},\
  }\bibfield  {title} {\bibinfo {title} {Ultracold atoms out of equilibrium},\
  }\href {https://doi.org/10.1146/annurev-conmatphys-031214-014548} {\bibfield
  {journal} {\bibinfo  {journal} {Annu. Rev. Condens. Matter Phys.}\ }\textbf
  {\bibinfo {volume} {6}},\ \bibinfo {pages} {201} (\bibinfo {year}
  {2015})}\BibitemShut {NoStop}%
\bibitem [{\citenamefont {Gross}\ and\ \citenamefont
  {Bloch}(2017)}]{Gross2017}%
  \BibitemOpen
  \bibfield  {author} {\bibinfo {author} {\bibfnamefont {C.}~\bibnamefont
  {Gross}}\ and\ \bibinfo {author} {\bibfnamefont {I.}~\bibnamefont {Bloch}},\
  }\bibfield  {title} {\bibinfo {title} {Quantum simulations with ultracold
  atoms in optical lattices},\ }\href {https://doi.org/10.1126/science.aal3837}
  {\bibfield  {journal} {\bibinfo  {journal} {Science}\ }\textbf {\bibinfo
  {volume} {357}},\ \bibinfo {pages} {995} (\bibinfo {year}
  {2017})}\BibitemShut {NoStop}%
\bibitem [{\citenamefont {Heyl}\ \emph {et~al.}(2013)\citenamefont {Heyl},
  \citenamefont {Polkovnikov},\ and\ \citenamefont {Kehrein}}]{heyl2013}%
  \BibitemOpen
  \bibfield  {author} {\bibinfo {author} {\bibfnamefont {M.}~\bibnamefont
  {Heyl}}, \bibinfo {author} {\bibfnamefont {A.}~\bibnamefont {Polkovnikov}},\
  and\ \bibinfo {author} {\bibfnamefont {S.}~\bibnamefont {Kehrein}},\
  }\bibfield  {title} {\bibinfo {title} {{Dynamical quantum phase transitions
  in the transverse-field Ising model}},\ }\href
  {https://doi.org/10.1103/PhysRevLett.110.135704} {\bibfield  {journal}
  {\bibinfo  {journal} {Phys. Rev. Lett.}\ }\textbf {\bibinfo {volume} {110}},\
  \bibinfo {pages} {135704} (\bibinfo {year} {2013})}\BibitemShut {NoStop}%
\bibitem [{\citenamefont {Heyl}(2018)}]{heyl2018}%
  \BibitemOpen
  \bibfield  {author} {\bibinfo {author} {\bibfnamefont {M.}~\bibnamefont
  {Heyl}},\ }\bibfield  {title} {\bibinfo {title} {Dynamical quantum phase
  transitions: a review},\ }\href {https://doi.org/10.1088/1361-6633/aaaf9a}
  {\bibfield  {journal} {\bibinfo  {journal} {Rep. Prog. Phys.}\ }\textbf
  {\bibinfo {volume} {81}},\ \bibinfo {pages} {054001} (\bibinfo {year}
  {2018})}\BibitemShut {NoStop}%
\bibitem [{\citenamefont {Karrasch}\ and\ \citenamefont
  {Schuricht}(2013)}]{KarraschSchuricht2013}%
  \BibitemOpen
  \bibfield  {author} {\bibinfo {author} {\bibfnamefont {C.}~\bibnamefont
  {Karrasch}}\ and\ \bibinfo {author} {\bibfnamefont {D.}~\bibnamefont
  {Schuricht}},\ }\bibfield  {title} {\bibinfo {title} {{Dynamical phase
  transitions after quenches in nonintegrable models}},\ }\href
  {https://doi.org/10.1103/PhysRevB.87.195104} {\bibfield  {journal} {\bibinfo
  {journal} {Phys. Rev. B}\ }\textbf {\bibinfo {volume} {87}},\ \bibinfo
  {pages} {195104} (\bibinfo {year} {2013})}\BibitemShut {NoStop}%
\bibitem [{\citenamefont {Vajna}\ and\ \citenamefont
  {D\'ora}(2014)}]{vajnaDora2014}%
  \BibitemOpen
  \bibfield  {author} {\bibinfo {author} {\bibfnamefont {S.}~\bibnamefont
  {Vajna}}\ and\ \bibinfo {author} {\bibfnamefont {B.}~\bibnamefont {D\'ora}},\
  }\bibfield  {title} {\bibinfo {title} {Disentangling dynamical phase
  transitions from equilibrium phase transitions},\ }\href
  {https://doi.org/10.1103/PhysRevB.89.161105} {\bibfield  {journal} {\bibinfo
  {journal} {Phys. Rev. B}\ }\textbf {\bibinfo {volume} {89}},\ \bibinfo
  {pages} {161105} (\bibinfo {year} {2014})}\BibitemShut {NoStop}%
\bibitem [{\citenamefont {Andraschko}\ and\ \citenamefont
  {Sirker}(2014)}]{andraschkoSirker2014}%
  \BibitemOpen
  \bibfield  {author} {\bibinfo {author} {\bibfnamefont {F.}~\bibnamefont
  {Andraschko}}\ and\ \bibinfo {author} {\bibfnamefont {J.}~\bibnamefont
  {Sirker}},\ }\bibfield  {title} {\bibinfo {title} {{Dynamical quantum phase
  transitions and the Loschmidt echo: A transfer matrix approach}},\ }\href
  {https://doi.org/10.1103/PhysRevB.89.125120} {\bibfield  {journal} {\bibinfo
  {journal} {Phys. Rev. B}\ }\textbf {\bibinfo {volume} {89}},\ \bibinfo
  {pages} {125120} (\bibinfo {year} {2014})}\BibitemShut {NoStop}%
\bibitem [{\citenamefont {Canovi}\ \emph
  {et~al.}(2014{\natexlab{a}})\citenamefont {Canovi}, \citenamefont {Werner},\
  and\ \citenamefont {Eckstein}}]{canovi2014}%
  \BibitemOpen
  \bibfield  {author} {\bibinfo {author} {\bibfnamefont {E.}~\bibnamefont
  {Canovi}}, \bibinfo {author} {\bibfnamefont {P.}~\bibnamefont {Werner}},\
  and\ \bibinfo {author} {\bibfnamefont {M.}~\bibnamefont {Eckstein}},\
  }\bibfield  {title} {\bibinfo {title} {First-order dynamical phase
  transitions},\ }\href {https://doi.org/10.1103/PhysRevLett.113.265702}
  {\bibfield  {journal} {\bibinfo  {journal} {Phys. Rev. Lett.}\ }\textbf
  {\bibinfo {volume} {113}},\ \bibinfo {pages} {265702} (\bibinfo {year}
  {2014}{\natexlab{a}})}\BibitemShut {NoStop}%
\bibitem [{\citenamefont {Canovi}\ \emph
  {et~al.}(2014{\natexlab{b}})\citenamefont {Canovi}, \citenamefont
  {Ercolessi}, \citenamefont {Naldesi}, \citenamefont {Taddia},\ and\
  \citenamefont {Vodola}}]{CanoviPRB2014}%
  \BibitemOpen
  \bibfield  {author} {\bibinfo {author} {\bibfnamefont {E.}~\bibnamefont
  {Canovi}}, \bibinfo {author} {\bibfnamefont {E.}~\bibnamefont {Ercolessi}},
  \bibinfo {author} {\bibfnamefont {P.}~\bibnamefont {Naldesi}}, \bibinfo
  {author} {\bibfnamefont {L.}~\bibnamefont {Taddia}},\ and\ \bibinfo {author}
  {\bibfnamefont {D.}~\bibnamefont {Vodola}},\ }\bibfield  {title} {\bibinfo
  {title} {Dynamics of entanglement entropy and entanglement spectrum crossing
  a quantum phase transition},\ }\href
  {https://doi.org/10.1103/PhysRevB.89.104303} {\bibfield  {journal} {\bibinfo
  {journal} {Phys. Rev. B}\ }\textbf {\bibinfo {volume} {89}},\ \bibinfo
  {pages} {104303} (\bibinfo {year} {2014}{\natexlab{b}})}\BibitemShut
  {NoStop}%
\bibitem [{\citenamefont {Torlai}\ \emph {et~al.}(2014)\citenamefont {Torlai},
  \citenamefont {Tagliacozzo},\ and\ \citenamefont {Chiara}}]{Torlai2014}%
  \BibitemOpen
  \bibfield  {author} {\bibinfo {author} {\bibfnamefont {G.}~\bibnamefont
  {Torlai}}, \bibinfo {author} {\bibfnamefont {L.}~\bibnamefont
  {Tagliacozzo}},\ and\ \bibinfo {author} {\bibfnamefont {G.~D.}\ \bibnamefont
  {Chiara}},\ }\bibfield  {title} {\bibinfo {title} {Dynamics of the
  entanglement spectrum in spin chains},\ }\href
  {https://doi.org/10.1088/1742-5468/2014/06/p06001} {\bibfield  {journal}
  {\bibinfo  {journal} {J. Stat. Mech.: Theory Exp.}\ }\textbf {\bibinfo
  {volume} {2014}}\bibinfo  {number} { (6)},\ \bibinfo {pages}
  {P06001}}\BibitemShut {NoStop}%
\bibitem [{\citenamefont {Heyl}(2014)}]{heyl2014}%
  \BibitemOpen
\bibfield  {number} {  }\bibfield  {author} {\bibinfo {author} {\bibfnamefont
  {M.}~\bibnamefont {Heyl}},\ }\bibfield  {title} {\bibinfo {title} {Dynamical
  quantum phase transitions in systems with broken-symmetry phases},\ }\href
  {https://doi.org/10.1103/PhysRevLett.113.205701} {\bibfield  {journal}
  {\bibinfo  {journal} {Phys. Rev. Lett.}\ }\textbf {\bibinfo {volume} {113}},\
  \bibinfo {pages} {205701} (\bibinfo {year} {2014})}\BibitemShut {NoStop}%
\bibitem [{\citenamefont {Heyl}(2015)}]{heyl2015}%
  \BibitemOpen
  \bibfield  {author} {\bibinfo {author} {\bibfnamefont {M.}~\bibnamefont
  {Heyl}},\ }\bibfield  {title} {\bibinfo {title} {Scaling and universality at
  dynamical quantum phase transitions},\ }\href
  {https://doi.org/10.1103/PhysRevLett.115.140602} {\bibfield  {journal}
  {\bibinfo  {journal} {Phys. Rev. Lett.}\ }\textbf {\bibinfo {volume} {115}},\
  \bibinfo {pages} {140602} (\bibinfo {year} {2015})}\BibitemShut {NoStop}%
\bibitem [{\citenamefont {Vajna}\ and\ \citenamefont
  {D\'ora}(2015)}]{vajnaDora2015}%
  \BibitemOpen
  \bibfield  {author} {\bibinfo {author} {\bibfnamefont {S.}~\bibnamefont
  {Vajna}}\ and\ \bibinfo {author} {\bibfnamefont {B.}~\bibnamefont {D\'ora}},\
  }\bibfield  {title} {\bibinfo {title} {Topological classification of
  dynamical phase transitions},\ }\href
  {https://doi.org/10.1103/PhysRevB.91.155127} {\bibfield  {journal} {\bibinfo
  {journal} {Phys. Rev. B}\ }\textbf {\bibinfo {volume} {91}},\ \bibinfo
  {pages} {155127} (\bibinfo {year} {2015})}\BibitemShut {NoStop}%
\bibitem [{\citenamefont {Sharma}\ \emph {et~al.}(2015)\citenamefont {Sharma},
  \citenamefont {Suzuki},\ and\ \citenamefont {Dutta}}]{Sharma2015}%
  \BibitemOpen
  \bibfield  {author} {\bibinfo {author} {\bibfnamefont {S.}~\bibnamefont
  {Sharma}}, \bibinfo {author} {\bibfnamefont {S.}~\bibnamefont {Suzuki}},\
  and\ \bibinfo {author} {\bibfnamefont {A.}~\bibnamefont {Dutta}},\ }\bibfield
   {title} {\bibinfo {title} {{Quenches and dynamical phase transitions in a
  nonintegrable quantum Ising model}},\ }\href
  {https://doi.org/10.1103/PhysRevB.92.104306} {\bibfield  {journal} {\bibinfo
  {journal} {Phys. Rev. B}\ }\textbf {\bibinfo {volume} {92}},\ \bibinfo
  {pages} {104306} (\bibinfo {year} {2015})}\BibitemShut {NoStop}%
\bibitem [{\citenamefont {Schmitt}\ and\ \citenamefont
  {Kehrein}(2015)}]{schmittKehrein2015}%
  \BibitemOpen
  \bibfield  {author} {\bibinfo {author} {\bibfnamefont {M.}~\bibnamefont
  {Schmitt}}\ and\ \bibinfo {author} {\bibfnamefont {S.}~\bibnamefont
  {Kehrein}},\ }\bibfield  {title} {\bibinfo {title} {{Dynamical quantum phase
  transitions in the Kitaev honeycomb model}},\ }\href
  {https://doi.org/10.1103/PhysRevB.92.075114} {\bibfield  {journal} {\bibinfo
  {journal} {Phys. Rev. B}\ }\textbf {\bibinfo {volume} {92}},\ \bibinfo
  {pages} {075114} (\bibinfo {year} {2015})}\BibitemShut {NoStop}%
\bibitem [{\citenamefont {Halimeh}\ and\ \citenamefont
  {Zauner-Stauber}(2017)}]{Halimeh2017}%
  \BibitemOpen
  \bibfield  {author} {\bibinfo {author} {\bibfnamefont {J.~C.}\ \bibnamefont
  {Halimeh}}\ and\ \bibinfo {author} {\bibfnamefont {V.}~\bibnamefont
  {Zauner-Stauber}},\ }\bibfield  {title} {\bibinfo {title} {Dynamical phase
  diagram of quantum spin chains with long-range interactions},\ }\href
  {https://doi.org/10.1103/PhysRevB.96.134427} {\bibfield  {journal} {\bibinfo
  {journal} {Phys. Rev. B}\ }\textbf {\bibinfo {volume} {96}},\ \bibinfo
  {pages} {134427} (\bibinfo {year} {2017})}\BibitemShut {NoStop}%
\bibitem [{\citenamefont {Weidinger}\ \emph {et~al.}(2017)\citenamefont
  {Weidinger}, \citenamefont {Heyl}, \citenamefont {Silva},\ and\ \citenamefont
  {Knap}}]{weidinger2017}%
  \BibitemOpen
  \bibfield  {author} {\bibinfo {author} {\bibfnamefont {S.~A.}\ \bibnamefont
  {Weidinger}}, \bibinfo {author} {\bibfnamefont {M.}~\bibnamefont {Heyl}},
  \bibinfo {author} {\bibfnamefont {A.}~\bibnamefont {Silva}},\ and\ \bibinfo
  {author} {\bibfnamefont {M.}~\bibnamefont {Knap}},\ }\bibfield  {title}
  {\bibinfo {title} {Dynamical quantum phase transitions in systems with
  continuous symmetry breaking},\ }\href
  {https://doi.org/10.1103/PhysRevB.96.134313} {\bibfield  {journal} {\bibinfo
  {journal} {Phys. Rev. B}\ }\textbf {\bibinfo {volume} {96}},\ \bibinfo
  {pages} {134313} (\bibinfo {year} {2017})}\BibitemShut {NoStop}%
\bibitem [{\citenamefont {Karrasch}\ and\ \citenamefont
  {Schuricht}(2017)}]{Karrasch2017}%
  \BibitemOpen
  \bibfield  {author} {\bibinfo {author} {\bibfnamefont {C.}~\bibnamefont
  {Karrasch}}\ and\ \bibinfo {author} {\bibfnamefont {D.}~\bibnamefont
  {Schuricht}},\ }\bibfield  {title} {\bibinfo {title} {{Dynamical quantum
  phase transitions in the quantum Potts chain}},\ }\href
  {https://doi.org/10.1103/PhysRevB.95.075143} {\bibfield  {journal} {\bibinfo
  {journal} {Phys. Rev. B}\ }\textbf {\bibinfo {volume} {95}},\ \bibinfo
  {pages} {075143} (\bibinfo {year} {2017})}\BibitemShut {NoStop}%
\bibitem [{\citenamefont {Homrighausen}\ \emph {et~al.}(2017)\citenamefont
  {Homrighausen}, \citenamefont {Abeling}, \citenamefont {Zauner-Stauber},\
  and\ \citenamefont {Halimeh}}]{Homrighausen2017}%
  \BibitemOpen
  \bibfield  {author} {\bibinfo {author} {\bibfnamefont {I.}~\bibnamefont
  {Homrighausen}}, \bibinfo {author} {\bibfnamefont {N.~O.}\ \bibnamefont
  {Abeling}}, \bibinfo {author} {\bibfnamefont {V.}~\bibnamefont
  {Zauner-Stauber}},\ and\ \bibinfo {author} {\bibfnamefont {J.~C.}\
  \bibnamefont {Halimeh}},\ }\bibfield  {title} {\bibinfo {title} {Anomalous
  dynamical phase in quantum spin chains with long-range interactions},\ }\href
  {https://doi.org/10.1103/PhysRevB.96.104436} {\bibfield  {journal} {\bibinfo
  {journal} {Phys. Rev. B}\ }\textbf {\bibinfo {volume} {96}},\ \bibinfo
  {pages} {104436} (\bibinfo {year} {2017})}\BibitemShut {NoStop}%
\bibitem [{\citenamefont {\ifmmode \check{Z}\else
  \v{Z}\fi{}unkovi\ifmmode~\check{c}\else \v{c}\fi{}}\ \emph
  {et~al.}(2018)\citenamefont {\ifmmode \check{Z}\else
  \v{Z}\fi{}unkovi\ifmmode~\check{c}\else \v{c}\fi{}}, \citenamefont {Heyl},
  \citenamefont {Knap},\ and\ \citenamefont {Silva}}]{zunkovic2018}%
  \BibitemOpen
  \bibfield  {author} {\bibinfo {author} {\bibfnamefont {B.}~\bibnamefont
  {\ifmmode \check{Z}\else \v{Z}\fi{}unkovi\ifmmode~\check{c}\else
  \v{c}\fi{}}}, \bibinfo {author} {\bibfnamefont {M.}~\bibnamefont {Heyl}},
  \bibinfo {author} {\bibfnamefont {M.}~\bibnamefont {Knap}},\ and\ \bibinfo
  {author} {\bibfnamefont {A.}~\bibnamefont {Silva}},\ }\bibfield  {title}
  {\bibinfo {title} {Dynamical quantum phase transitions in spin chains with
  long-range interactions: Merging different concepts of nonequilibrium
  criticality},\ }\href {https://doi.org/10.1103/PhysRevLett.120.130601}
  {\bibfield  {journal} {\bibinfo  {journal} {Phys. Rev. Lett.}\ }\textbf
  {\bibinfo {volume} {120}},\ \bibinfo {pages} {130601} (\bibinfo {year}
  {2018})}\BibitemShut {NoStop}%
\bibitem [{\citenamefont {Schmitt}\ and\ \citenamefont
  {Heyl}(2018)}]{schmittHeyl2018}%
  \BibitemOpen
  \bibfield  {author} {\bibinfo {author} {\bibfnamefont {M.}~\bibnamefont
  {Schmitt}}\ and\ \bibinfo {author} {\bibfnamefont {M.}~\bibnamefont {Heyl}},\
  }\bibfield  {title} {\bibinfo {title} {{Quantum dynamics in transverse-field
  Ising models from classical networks}},\ }\href
  {https://doi.org/10.21468/SciPostPhys.4.2.013} {\bibfield  {journal}
  {\bibinfo  {journal} {SciPost Phys.}\ }\textbf {\bibinfo {volume} {4}},\
  \bibinfo {pages} {013} (\bibinfo {year} {2018})}\BibitemShut {NoStop}%
\bibitem [{\citenamefont {Trapin}\ and\ \citenamefont
  {Heyl}(2018)}]{Trapin2018}%
  \BibitemOpen
  \bibfield  {author} {\bibinfo {author} {\bibfnamefont {D.}~\bibnamefont
  {Trapin}}\ and\ \bibinfo {author} {\bibfnamefont {M.}~\bibnamefont {Heyl}},\
  }\bibfield  {title} {\bibinfo {title} {{Constructing effective free energies
  for dynamical quantum phase transitions in the transverse-field Ising
  chain}},\ }\href {https://doi.org/10.1103/PhysRevB.97.174303} {\bibfield
  {journal} {\bibinfo  {journal} {Phys. Rev. B}\ }\textbf {\bibinfo {volume}
  {97}},\ \bibinfo {pages} {174303} (\bibinfo {year} {2018})}\BibitemShut
  {NoStop}%
\bibitem [{\citenamefont {Gurarie}(2019)}]{Gurarie2019}%
  \BibitemOpen
  \bibfield  {author} {\bibinfo {author} {\bibfnamefont {V.}~\bibnamefont
  {Gurarie}},\ }\bibfield  {title} {\bibinfo {title} {{Dynamical quantum phase
  transitions in the random field Ising model}},\ }\href
  {https://doi.org/10.1103/PhysRevA.100.031601} {\bibfield  {journal} {\bibinfo
   {journal} {Phys. Rev. A}\ }\textbf {\bibinfo {volume} {100}},\ \bibinfo
  {pages} {031601} (\bibinfo {year} {2019})}\BibitemShut {NoStop}%
\bibitem [{\citenamefont {De~Nicola}\ \emph {et~al.}(2019)\citenamefont
  {De~Nicola}, \citenamefont {Doyon},\ and\ \citenamefont
  {Bhaseen}}]{DeNicola2019}%
  \BibitemOpen
  \bibfield  {author} {\bibinfo {author} {\bibfnamefont {S.}~\bibnamefont
  {De~Nicola}}, \bibinfo {author} {\bibfnamefont {B.}~\bibnamefont {Doyon}},\
  and\ \bibinfo {author} {\bibfnamefont {M.~J.}\ \bibnamefont {Bhaseen}},\
  }\bibfield  {title} {\bibinfo {title} {Stochastic approach to non-equilibrium
  quantum spin systems},\ }\href {https://doi.org/10.1088/1751-8121/aaf9be}
  {\bibfield  {journal} {\bibinfo  {journal} {J. Phys. A: Math. Theor.}\
  }\textbf {\bibinfo {volume} {52}},\ \bibinfo {pages} {05LT02} (\bibinfo
  {year} {2019})}\BibitemShut {NoStop}%
\bibitem [{\citenamefont {Huang}\ \emph {et~al.}(2019)\citenamefont {Huang},
  \citenamefont {Banerjee},\ and\ \citenamefont {Heyl}}]{Huang2019}%
  \BibitemOpen
  \bibfield  {author} {\bibinfo {author} {\bibfnamefont {Y.-P.}\ \bibnamefont
  {Huang}}, \bibinfo {author} {\bibfnamefont {D.}~\bibnamefont {Banerjee}},\
  and\ \bibinfo {author} {\bibfnamefont {M.}~\bibnamefont {Heyl}},\ }\bibfield
  {title} {\bibinfo {title} {Dynamical quantum phase transitions in {U(1)}
  quantum link models},\ }\href
  {https://doi.org/10.1103/PhysRevLett.122.250401} {\bibfield  {journal}
  {\bibinfo  {journal} {Phys. Rev. Lett.}\ }\textbf {\bibinfo {volume} {122}},\
  \bibinfo {pages} {250401} (\bibinfo {year} {2019})}\BibitemShut {NoStop}%
\bibitem [{\citenamefont {Lacki}\ and\ \citenamefont {Heyl}(2019)}]{Lacki2019}%
  \BibitemOpen
  \bibfield  {author} {\bibinfo {author} {\bibfnamefont {M.}~\bibnamefont
  {Lacki}}\ and\ \bibinfo {author} {\bibfnamefont {M.}~\bibnamefont {Heyl}},\
  }\bibfield  {title} {\bibinfo {title} {Dynamical quantum phase transitions in
  collapse and revival oscillations of a quenched superfluid},\ }\href
  {https://doi.org/10.1103/PhysRevB.99.121107} {\bibfield  {journal} {\bibinfo
  {journal} {Phys. Rev. B}\ }\textbf {\bibinfo {volume} {99}},\ \bibinfo
  {pages} {121107} (\bibinfo {year} {2019})}\BibitemShut {NoStop}%
\bibitem [{\citenamefont {Jafari}(2019)}]{jafari2019}%
  \BibitemOpen
  \bibfield  {author} {\bibinfo {author} {\bibfnamefont {R.}~\bibnamefont
  {Jafari}},\ }\bibfield  {title} {\bibinfo {title} {Dynamical quantum phase
  transition and quasi particle excitation},\ }\href
  {https://doi.org/10.1038/s41598-019-39595-3} {\bibfield  {journal} {\bibinfo
  {journal} {Sci. Rep.}\ }\textbf {\bibinfo {volume} {9}},\ \bibinfo {pages}
  {2871} (\bibinfo {year} {2019})}\BibitemShut {NoStop}%
\bibitem [{\citenamefont {Heyl}(2019)}]{heyl2019}%
  \BibitemOpen
  \bibfield  {author} {\bibinfo {author} {\bibfnamefont {M.}~\bibnamefont
  {Heyl}},\ }\bibfield  {title} {\bibinfo {title} {Dynamical quantum phase
  transitions: A brief survey},\ }\href
  {https://doi.org/10.1209/0295-5075/125/26001} {\bibfield  {journal} {\bibinfo
   {journal} {{EPL}}\ }\textbf {\bibinfo {volume} {125}},\ \bibinfo {pages}
  {26001} (\bibinfo {year} {2019})}\BibitemShut {NoStop}%
\bibitem [{\citenamefont {Jurcevic}\ \emph {et~al.}(2017)\citenamefont
  {Jurcevic}, \citenamefont {Shen}, \citenamefont {Hauke}, \citenamefont
  {Maier}, \citenamefont {Brydges}, \citenamefont {Hempel}, \citenamefont
  {Lanyon}, \citenamefont {Heyl}, \citenamefont {Blatt},\ and\ \citenamefont
  {Roos}}]{jurcevic2017}%
  \BibitemOpen
  \bibfield  {author} {\bibinfo {author} {\bibfnamefont {P.}~\bibnamefont
  {Jurcevic}}, \bibinfo {author} {\bibfnamefont {H.}~\bibnamefont {Shen}},
  \bibinfo {author} {\bibfnamefont {P.}~\bibnamefont {Hauke}}, \bibinfo
  {author} {\bibfnamefont {C.}~\bibnamefont {Maier}}, \bibinfo {author}
  {\bibfnamefont {T.}~\bibnamefont {Brydges}}, \bibinfo {author} {\bibfnamefont
  {C.}~\bibnamefont {Hempel}}, \bibinfo {author} {\bibfnamefont {B.~P.}\
  \bibnamefont {Lanyon}}, \bibinfo {author} {\bibfnamefont {M.}~\bibnamefont
  {Heyl}}, \bibinfo {author} {\bibfnamefont {R.}~\bibnamefont {Blatt}},\ and\
  \bibinfo {author} {\bibfnamefont {C.~F.}\ \bibnamefont {Roos}},\ }\bibfield
  {title} {\bibinfo {title} {Direct observation of dynamical quantum phase
  transitions in an interacting many-body system},\ }\href
  {https://doi.org/10.1103/PhysRevLett.119.080501} {\bibfield  {journal}
  {\bibinfo  {journal} {Phys. Rev. Lett.}\ }\textbf {\bibinfo {volume} {119}},\
  \bibinfo {pages} {080501} (\bibinfo {year} {2017})}\BibitemShut {NoStop}%
\bibitem [{\citenamefont {Guo}\ \emph {et~al.}(2019)\citenamefont {Guo},
  \citenamefont {Yang}, \citenamefont {Zeng}, \citenamefont {Peng},
  \citenamefont {Li}, \citenamefont {Deng}, \citenamefont {Jin}, \citenamefont
  {Chen}, \citenamefont {Zheng},\ and\ \citenamefont {Fan}}]{Guo2019}%
  \BibitemOpen
  \bibfield  {author} {\bibinfo {author} {\bibfnamefont {X.-Y.}\ \bibnamefont
  {Guo}}, \bibinfo {author} {\bibfnamefont {C.}~\bibnamefont {Yang}}, \bibinfo
  {author} {\bibfnamefont {Y.}~\bibnamefont {Zeng}}, \bibinfo {author}
  {\bibfnamefont {Y.}~\bibnamefont {Peng}}, \bibinfo {author} {\bibfnamefont
  {H.-K.}\ \bibnamefont {Li}}, \bibinfo {author} {\bibfnamefont
  {H.}~\bibnamefont {Deng}}, \bibinfo {author} {\bibfnamefont {Y.-R.}\
  \bibnamefont {Jin}}, \bibinfo {author} {\bibfnamefont {S.}~\bibnamefont
  {Chen}}, \bibinfo {author} {\bibfnamefont {D.}~\bibnamefont {Zheng}},\ and\
  \bibinfo {author} {\bibfnamefont {H.}~\bibnamefont {Fan}},\ }\bibfield
  {title} {\bibinfo {title} {Observation of a dynamical quantum phase
  transition by a superconducting qubit simulation},\ }\href
  {https://doi.org/10.1103/PhysRevApplied.11.044080} {\bibfield  {journal}
  {\bibinfo  {journal} {Phys. Rev. Applied}\ }\textbf {\bibinfo {volume}
  {11}},\ \bibinfo {pages} {044080} (\bibinfo {year} {2019})}\BibitemShut
  {NoStop}%
\bibitem [{\citenamefont {Fl{\"{a}}schner}\ \emph {et~al.}(2018)\citenamefont
  {Fl{\"{a}}schner}, \citenamefont {Vogel}, \citenamefont {Tarnowski},
  \citenamefont {Rem}, \citenamefont {L{\"{u}}hmann}, \citenamefont {Heyl},
  \citenamefont {Budich}, \citenamefont {Mathey}, \citenamefont {Sengstock},\
  and\ \citenamefont {Weitenberg}}]{Flaschner2018}%
  \BibitemOpen
  \bibfield  {author} {\bibinfo {author} {\bibfnamefont {N.}~\bibnamefont
  {Fl{\"{a}}schner}}, \bibinfo {author} {\bibfnamefont {D.}~\bibnamefont
  {Vogel}}, \bibinfo {author} {\bibfnamefont {M.}~\bibnamefont {Tarnowski}},
  \bibinfo {author} {\bibfnamefont {B.~S.}\ \bibnamefont {Rem}}, \bibinfo
  {author} {\bibfnamefont {D.-S.}\ \bibnamefont {L{\"{u}}hmann}}, \bibinfo
  {author} {\bibfnamefont {M.}~\bibnamefont {Heyl}}, \bibinfo {author}
  {\bibfnamefont {J.~C.}\ \bibnamefont {Budich}}, \bibinfo {author}
  {\bibfnamefont {L.}~\bibnamefont {Mathey}}, \bibinfo {author} {\bibfnamefont
  {K.}~\bibnamefont {Sengstock}},\ and\ \bibinfo {author} {\bibfnamefont
  {C.}~\bibnamefont {Weitenberg}},\ }\bibfield  {title} {\bibinfo {title}
  {{Observation of dynamical vortices after quenches in a system with
  topology}},\ }\href {https://doi.org/10.1038/s41567-017-0013-8} {\bibfield
  {journal} {\bibinfo  {journal} {Nat. Phys.}\ }\textbf {\bibinfo {volume}
  {14}},\ \bibinfo {pages} {265} (\bibinfo {year} {2018})}\BibitemShut
  {NoStop}%
\bibitem [{\citenamefont {Tian}\ \emph {et~al.}(2019)\citenamefont {Tian},
  \citenamefont {Ke}, \citenamefont {Zhang}, \citenamefont {Lin}, \citenamefont
  {Shi}, \citenamefont {Huang}, \citenamefont {Lee},\ and\ \citenamefont
  {Du}}]{Tian2018}%
  \BibitemOpen
  \bibfield  {author} {\bibinfo {author} {\bibfnamefont {T.}~\bibnamefont
  {Tian}}, \bibinfo {author} {\bibfnamefont {Y.}~\bibnamefont {Ke}}, \bibinfo
  {author} {\bibfnamefont {L.}~\bibnamefont {Zhang}}, \bibinfo {author}
  {\bibfnamefont {S.}~\bibnamefont {Lin}}, \bibinfo {author} {\bibfnamefont
  {Z.}~\bibnamefont {Shi}}, \bibinfo {author} {\bibfnamefont {P.}~\bibnamefont
  {Huang}}, \bibinfo {author} {\bibfnamefont {C.}~\bibnamefont {Lee}},\ and\
  \bibinfo {author} {\bibfnamefont {J.}~\bibnamefont {Du}},\ }\bibfield
  {title} {\bibinfo {title} {Observation of dynamical phase transitions in a
  topological nanomechanical system},\ }\href
  {https://doi.org/10.1103/PhysRevB.100.024310} {\bibfield  {journal} {\bibinfo
   {journal} {Phys. Rev. B}\ }\textbf {\bibinfo {volume} {100}},\ \bibinfo
  {pages} {024310} (\bibinfo {year} {2019})}\BibitemShut {NoStop}%
\bibitem [{\citenamefont {Wang}\ \emph {et~al.}(2019)\citenamefont {Wang},
  \citenamefont {Qiu}, \citenamefont {Xiao}, \citenamefont {Zhan},
  \citenamefont {Bian}, \citenamefont {Yi},\ and\ \citenamefont
  {Xue}}]{Wang2019}%
  \BibitemOpen
  \bibfield  {author} {\bibinfo {author} {\bibfnamefont {K.}~\bibnamefont
  {Wang}}, \bibinfo {author} {\bibfnamefont {X.}~\bibnamefont {Qiu}}, \bibinfo
  {author} {\bibfnamefont {L.}~\bibnamefont {Xiao}}, \bibinfo {author}
  {\bibfnamefont {X.}~\bibnamefont {Zhan}}, \bibinfo {author} {\bibfnamefont
  {Z.}~\bibnamefont {Bian}}, \bibinfo {author} {\bibfnamefont {W.}~\bibnamefont
  {Yi}},\ and\ \bibinfo {author} {\bibfnamefont {P.}~\bibnamefont {Xue}},\
  }\bibfield  {title} {\bibinfo {title} {Simulating dynamic quantum phase
  transitions in photonic quantum walks},\ }\href
  {https://doi.org/10.1103/PhysRevLett.122.020501} {\bibfield  {journal}
  {\bibinfo  {journal} {Phys. Rev. Lett.}\ }\textbf {\bibinfo {volume} {122}},\
  \bibinfo {pages} {020501} (\bibinfo {year} {2019})}\BibitemShut {NoStop}%
\bibitem [{\citenamefont {Xu}\ \emph {et~al.}(2020)\citenamefont {Xu},
  \citenamefont {Wang}, \citenamefont {Heyl}, \citenamefont {Budich},
  \citenamefont {Pan}, \citenamefont {Chen}, \citenamefont {Jan}, \citenamefont
  {Sun}, \citenamefont {Xu}, \citenamefont {Han}, \citenamefont {Li},\ and\
  \citenamefont {Guo}}]{Xu2020}%
  \BibitemOpen
  \bibfield  {author} {\bibinfo {author} {\bibfnamefont {X.-Y.}\ \bibnamefont
  {Xu}}, \bibinfo {author} {\bibfnamefont {Q.-Q.}\ \bibnamefont {Wang}},
  \bibinfo {author} {\bibfnamefont {M.}~\bibnamefont {Heyl}}, \bibinfo {author}
  {\bibfnamefont {J.~C.}\ \bibnamefont {Budich}}, \bibinfo {author}
  {\bibfnamefont {W.-W.}\ \bibnamefont {Pan}}, \bibinfo {author} {\bibfnamefont
  {Z.}~\bibnamefont {Chen}}, \bibinfo {author} {\bibfnamefont {M.}~\bibnamefont
  {Jan}}, \bibinfo {author} {\bibfnamefont {K.}~\bibnamefont {Sun}}, \bibinfo
  {author} {\bibfnamefont {J.-S.}\ \bibnamefont {Xu}}, \bibinfo {author}
  {\bibfnamefont {Y.-J.}\ \bibnamefont {Han}}, \bibinfo {author} {\bibfnamefont
  {C.-F.}\ \bibnamefont {Li}},\ and\ \bibinfo {author} {\bibfnamefont {G.-C.}\
  \bibnamefont {Guo}},\ }\bibfield  {title} {\bibinfo {title} {Measuring a
  dynamical topological order parameter in quantum walks},\ }\href
  {https://doi.org/10.1038/s41377-019-0237-8} {\bibfield  {journal} {\bibinfo
  {journal} {Light Sci. Appl.}\ }\textbf {\bibinfo {volume} {9}},\ \bibinfo
  {pages} {7} (\bibinfo {year} {2020})}\BibitemShut {NoStop}%
\bibitem [{\citenamefont {Rylands}\ and\ \citenamefont
  {Galitski}(2020)}]{Rylands2020}%
  \BibitemOpen
  \bibfield  {author} {\bibinfo {author} {\bibfnamefont {C.}~\bibnamefont
  {Rylands}}\ and\ \bibinfo {author} {\bibfnamefont {V.}~\bibnamefont
  {Galitski}},\ }\bibfield  {title} {\bibinfo {title} {Dynamical quantum phase
  transitions and recurrences in the non-equilibrium {BCS} model},\ }\Eprint
  {https://arxiv.org/abs/2001.10084} {arXiv:2001.10084 [cond-mat.supr-con]}
  (\bibinfo {year} {2020})\BibitemShut {NoStop}%
\bibitem [{\citenamefont {Feldmeier}\ \emph {et~al.}(2019)\citenamefont
  {Feldmeier}, \citenamefont {Pollmann},\ and\ \citenamefont
  {Knap}}]{feldmeierPollmannKnap2019}%
  \BibitemOpen
  \bibfield  {author} {\bibinfo {author} {\bibfnamefont {J.}~\bibnamefont
  {Feldmeier}}, \bibinfo {author} {\bibfnamefont {F.}~\bibnamefont
  {Pollmann}},\ and\ \bibinfo {author} {\bibfnamefont {M.}~\bibnamefont
  {Knap}},\ }\bibfield  {title} {\bibinfo {title} {Emergent glassy dynamics in
  a quantum dimer model},\ }\href
  {https://doi.org/10.1103/PhysRevLett.123.040601} {\bibfield  {journal}
  {\bibinfo  {journal} {Phys. Rev. Lett.}\ }\textbf {\bibinfo {volume} {123}},\
  \bibinfo {pages} {040601} (\bibinfo {year} {2019})}\BibitemShut {NoStop}%
\bibitem [{\citenamefont {Halimeh}\ \emph {et~al.}(2020)\citenamefont
  {Halimeh}, \citenamefont {Trapin}, \citenamefont {Damme},\ and\ \citenamefont
  {Heyl}}]{halimeh2020local}%
  \BibitemOpen
  \bibfield  {author} {\bibinfo {author} {\bibfnamefont {J.~C.}\ \bibnamefont
  {Halimeh}}, \bibinfo {author} {\bibfnamefont {D.}~\bibnamefont {Trapin}},
  \bibinfo {author} {\bibfnamefont {M.~V.}\ \bibnamefont {Damme}},\ and\
  \bibinfo {author} {\bibfnamefont {M.}~\bibnamefont {Heyl}},\ }\href@noop {}
  {\bibinfo {title} {Local measures of dynamical quantum phase transitions}}
  (\bibinfo {year} {2020}),\ \Eprint {https://arxiv.org/abs/2010.07307}
  {arXiv:2010.07307 [cond-mat.quant-gas]} \BibitemShut {NoStop}%
\bibitem [{\citenamefont {Bandyopadhyay}\ \emph {et~al.}(2020)\citenamefont
  {Bandyopadhyay}, \citenamefont {Polkovnikov},\ and\ \citenamefont
  {Dutta}}]{bandyopadhyay2020observing}%
  \BibitemOpen
  \bibfield  {author} {\bibinfo {author} {\bibfnamefont {S.}~\bibnamefont
  {Bandyopadhyay}}, \bibinfo {author} {\bibfnamefont {A.}~\bibnamefont
  {Polkovnikov}},\ and\ \bibinfo {author} {\bibfnamefont {A.}~\bibnamefont
  {Dutta}},\ }\href@noop {} {\bibinfo {title} {Observing dynamical quantum
  phase transitions through quasi-local string operators}} (\bibinfo {year}
  {2020}),\ \Eprint {https://arxiv.org/abs/2011.03906} {arXiv:2011.03906
  [cond-mat.stat-mech]} \BibitemShut {NoStop}%
\bibitem [{\citenamefont {Surace}\ \emph {et~al.}(2020)\citenamefont {Surace},
  \citenamefont {Tagliacozzo},\ and\ \citenamefont {Tonni}}]{Surace2020}%
  \BibitemOpen
  \bibfield  {author} {\bibinfo {author} {\bibfnamefont {J.}~\bibnamefont
  {Surace}}, \bibinfo {author} {\bibfnamefont {L.}~\bibnamefont
  {Tagliacozzo}},\ and\ \bibinfo {author} {\bibfnamefont {E.}~\bibnamefont
  {Tonni}},\ }\bibfield  {title} {\bibinfo {title} {Operator content of
  entanglement spectra in the transverse field {Ising chain} after global
  quenches},\ }\href {https://doi.org/10.1103/PhysRevB.101.241107} {\bibfield
  {journal} {\bibinfo  {journal} {Phys. Rev. B}\ }\textbf {\bibinfo {volume}
  {101}},\ \bibinfo {pages} {241107} (\bibinfo {year} {2020})}\BibitemShut
  {NoStop}%
\bibitem [{\citenamefont {Paeckel}\ \emph {et~al.}(2019)\citenamefont
  {Paeckel}, \citenamefont {K{\"{o}}hler}, \citenamefont {Swoboda},
  \citenamefont {Manmana}, \citenamefont {Schollw{\"{o}}ck},\ and\
  \citenamefont {Hubig}}]{Paeckel2019}%
  \BibitemOpen
  \bibfield  {author} {\bibinfo {author} {\bibfnamefont {S.}~\bibnamefont
  {Paeckel}}, \bibinfo {author} {\bibfnamefont {T.}~\bibnamefont
  {K{\"{o}}hler}}, \bibinfo {author} {\bibfnamefont {A.}~\bibnamefont
  {Swoboda}}, \bibinfo {author} {\bibfnamefont {S.~R.}\ \bibnamefont
  {Manmana}}, \bibinfo {author} {\bibfnamefont {U.}~\bibnamefont
  {Schollw{\"{o}}ck}},\ and\ \bibinfo {author} {\bibfnamefont {C.}~\bibnamefont
  {Hubig}},\ }\bibfield  {title} {\bibinfo {title} {{Time-evolution methods for
  matrix-product states}},\ }\href
  {https://doi.org/https://doi.org/10.1016/j.aop.2019.167998} {\bibfield
  {journal} {\bibinfo  {journal} {Ann. Phys. (N. Y).}\ }\textbf {\bibinfo
  {volume} {411}},\ \bibinfo {pages} {167998} (\bibinfo {year}
  {2019})}\BibitemShut {NoStop}%
\bibitem [{\citenamefont {Lieb}\ and\ \citenamefont
  {Robinson}(1972)}]{Lieb1972}%
  \BibitemOpen
  \bibfield  {author} {\bibinfo {author} {\bibfnamefont {E.~H.}\ \bibnamefont
  {Lieb}}\ and\ \bibinfo {author} {\bibfnamefont {D.~W.}\ \bibnamefont
  {Robinson}},\ }\bibfield  {title} {\bibinfo {title} {The finite group
  velocity of quantum spin systems},\ }\href
  {https://doi.org/10.1007/BF01645779} {\bibfield  {journal} {\bibinfo
  {journal} {Comm. Math. Phys.}\ }\textbf {\bibinfo {volume} {28}},\ \bibinfo
  {pages} {251} (\bibinfo {year} {1972})}\BibitemShut {NoStop}%
\bibitem [{\citenamefont {Eisert}\ and\ \citenamefont
  {Osborne}(2006)}]{Eisert2006}%
  \BibitemOpen
  \bibfield  {author} {\bibinfo {author} {\bibfnamefont {J.}~\bibnamefont
  {Eisert}}\ and\ \bibinfo {author} {\bibfnamefont {T.~J.}\ \bibnamefont
  {Osborne}},\ }\bibfield  {title} {\bibinfo {title} {General entanglement
  scaling laws from time evolution},\ }\href
  {https://doi.org/10.1103/PhysRevLett.97.150404} {\bibfield  {journal}
  {\bibinfo  {journal} {Phys. Rev. Lett.}\ }\textbf {\bibinfo {volume} {97}},\
  \bibinfo {pages} {150404} (\bibinfo {year} {2006})}\BibitemShut {NoStop}%
\bibitem [{\citenamefont {Vidal}(2007)}]{Vidal2006}%
  \BibitemOpen
  \bibfield  {author} {\bibinfo {author} {\bibfnamefont {G.}~\bibnamefont
  {Vidal}},\ }\bibfield  {title} {\bibinfo {title} {Classical simulation of
  infinite-size quantum lattice systems in one spatial dimension},\ }\href
  {https://doi.org/10.1103/PhysRevLett.98.070201} {\bibfield  {journal}
  {\bibinfo  {journal} {Phys. Rev. Lett.}\ }\textbf {\bibinfo {volume} {98}},\
  \bibinfo {pages} {070201} (\bibinfo {year} {2007})}\BibitemShut {NoStop}%
\bibitem [{\citenamefont {Or\'us}\ and\ \citenamefont
  {Vidal}(2008)}]{Orus2008}%
  \BibitemOpen
  \bibfield  {author} {\bibinfo {author} {\bibfnamefont {R.}~\bibnamefont
  {Or\'us}}\ and\ \bibinfo {author} {\bibfnamefont {G.}~\bibnamefont {Vidal}},\
  }\bibfield  {title} {\bibinfo {title} {Infinite time-evolving block
  decimation algorithm beyond unitary evolution},\ }\href
  {https://doi.org/10.1103/PhysRevB.78.155117} {\bibfield  {journal} {\bibinfo
  {journal} {Phys. Rev. B}\ }\textbf {\bibinfo {volume} {78}},\ \bibinfo
  {pages} {155117} (\bibinfo {year} {2008})}\BibitemShut {NoStop}%
\bibitem [{\citenamefont {Piroli}\ \emph {et~al.}(2018)\citenamefont {Piroli},
  \citenamefont {Pozsgay},\ and\ \citenamefont {Vernier}}]{piroli2018}%
  \BibitemOpen
  \bibfield  {author} {\bibinfo {author} {\bibfnamefont {L.}~\bibnamefont
  {Piroli}}, \bibinfo {author} {\bibfnamefont {B.}~\bibnamefont {Pozsgay}},\
  and\ \bibinfo {author} {\bibfnamefont {E.}~\bibnamefont {Vernier}},\
  }\bibfield  {title} {\bibinfo {title} {{Non-analytic behavior of the
  Loschmidt echo in XXZ spin chains: Exact results}},\ }\href
  {https://doi.org/https://doi.org/10.1016/j.nuclphysb.2018.06.015} {\bibfield
  {journal} {\bibinfo  {journal} {Nucl. Phys. B.}\ }\textbf {\bibinfo {volume}
  {933}},\ \bibinfo {pages} {454 } (\bibinfo {year} {2018})}\BibitemShut
  {NoStop}%
\bibitem [{SOM()}]{SOM}%
  \BibitemOpen
  \href@noop {} {}\bibinfo {note} {{ See Supplemental Material, which cites Refs. \cite{Crosswhite2008} and \cite{mussardo2009}}}\BibitemShut
  {NoStop}%
  \bibitem [{\citenamefont {Crosswhite}\ and\ \citenamefont
  {Bacon}(2008)}]{Crosswhite2008}%
  \BibitemOpen
  \bibfield  {author} {\bibinfo {author} {\bibfnamefont {G.~M.}\ \bibnamefont
  {Crosswhite}}\ and\ \bibinfo {author} {\bibfnamefont {D.}~\bibnamefont
  {Bacon}},\ }\bibfield  {title} {\bibinfo {title} {Finite automata for caching
  in matrix product algorithms},\ }\href
  {https://doi.org/10.1103/PhysRevA.78.012356} {\bibfield  {journal} {\bibinfo
  {journal} {Phys. Rev. A}\ }\textbf {\bibinfo {volume} {78}},\ \bibinfo
  {pages} {012356} (\bibinfo {year} {2008})}\BibitemShut {NoStop}%
\bibitem [{\citenamefont {Mussardo}(2009)}]{mussardo2009}%
  \BibitemOpen
  \bibfield  {author} {\bibinfo {author} {\bibfnamefont {G.}~\bibnamefont
  {Mussardo}},\ }\href@noop {} {\emph {\bibinfo {title} {Statistical Field
  Theory: An Introduction to Exactly Solved Models in Statistical Physics}}},\
  Oxford Graduate Texts\ (\bibinfo  {publisher} {OUP Oxford},\ \bibinfo {year}
  {2009})\BibitemShut {NoStop}%
\bibitem [{\citenamefont {Calabrese}\ \emph {et~al.}(2012)\citenamefont
  {Calabrese}, \citenamefont {Essler},\ and\ \citenamefont
  {Fagotti}}]{Calabrese2012}%
  \BibitemOpen
  \bibfield  {author} {\bibinfo {author} {\bibfnamefont {P.}~\bibnamefont
  {Calabrese}}, \bibinfo {author} {\bibfnamefont {F.~H.~L.}\ \bibnamefont
  {Essler}},\ and\ \bibinfo {author} {\bibfnamefont {M.}~\bibnamefont
  {Fagotti}},\ }\bibfield  {title} {\bibinfo {title} {Quantum quench in the
  transverse field {Ising} chain: I. time evolution of order parameter
  correlators},\ }\href
  {https://doi.org/https://doi.org/10.1088/1742-5468/2012/07/P07016} {\bibfield
   {journal} {\bibinfo  {journal} {J. Stat. Mech.: Theory Exp.}\ }\textbf
  {\bibinfo {volume} {2012}}\bibinfo  {number} { (07)},\ \bibinfo {pages}
  {P07016}}\BibitemShut {NoStop}%
\bibitem [{\citenamefont {Trapin}\ \emph {et~al.}(2020)\citenamefont {Trapin},
  \citenamefont {Halimeh},\ and\ \citenamefont {Heyl}}]{Trapin2020}%
  \BibitemOpen
\bibfield  {number} {  }\bibfield  {author} {\bibinfo {author} {\bibfnamefont
  {D.}~\bibnamefont {Trapin}}, \bibinfo {author} {\bibfnamefont {J.~C.}\
  \bibnamefont {Halimeh}},\ and\ \bibinfo {author} {\bibfnamefont
  {M.}~\bibnamefont {Heyl}},\ }\bibfield  {title} {\bibinfo {title}
  {{Unconventional critical exponents at dynamical quantum phase transitions in
  a random Ising chain}},\ }\Eprint {https://arxiv.org/abs/2005.06481}
  {arXiv:2005.06481 [cond-mat.stat-mech]}  (\bibinfo {year} {2020})\BibitemShut
  {NoStop}%
\bibitem [{\citenamefont {{Bergholtz}}\ \emph {et~al.}(2019)\citenamefont
  {{Bergholtz}}, \citenamefont {{Budich}},\ and\ \citenamefont
  {{Kunst}}}]{Berg19}%
  \BibitemOpen
  \bibfield  {author} {\bibinfo {author} {\bibfnamefont {E.~J.}\ \bibnamefont
  {{Bergholtz}}}, \bibinfo {author} {\bibfnamefont {J.~C.}\ \bibnamefont
  {{Budich}}},\ and\ \bibinfo {author} {\bibfnamefont {F.~K.}\ \bibnamefont
  {{Kunst}}},\ }\bibfield  {title} {\bibinfo {title} {{Exceptional topology of
  non-Hermitian systems}},\ }\Eprint {https://arxiv.org/abs/1912.10048}
  {arXiv:1912.10048 [cond-mat.mes-hall]}  (\bibinfo {year} {2019})\BibitemShut
  {NoStop}%
\bibitem [{\citenamefont {Ashida}\ \emph {et~al.}(2020)\citenamefont {Ashida},
  \citenamefont {Gong},\ and\ \citenamefont {Ueda}}]{Ashida2020}%
  \BibitemOpen
  \bibfield  {author} {\bibinfo {author} {\bibfnamefont {Y.}~\bibnamefont
  {Ashida}}, \bibinfo {author} {\bibfnamefont {Z.}~\bibnamefont {Gong}},\ and\
  \bibinfo {author} {\bibfnamefont {M.}~\bibnamefont {Ueda}},\ }\href@noop {}
  {\bibinfo {title} {{Non-Hermitian physics}}} (\bibinfo {year} {2020}),\
  \Eprint {https://arxiv.org/abs/2006.01837} {arXiv:2006.01837
  [cond-mat.mes-hall]} \BibitemShut {NoStop}%
\end{thebibliography}
\end{document}